\documentclass[10pt,a4paper]{article}

\usepackage[utf8]{inputenc}
\usepackage[english]{babel}
\usepackage{amssymb,amsmath}
\usepackage{euscript}

\begin{document}
\large

\newpage
\begin{center}
\LARGE{\bf A Set-Matrix Duality Principle
\\for the Dirac Equation}
\end{center}
\vspace{0.1mm}
\begin{center}
{\bf Rasulkhozha S. Sharafiddinov}
\end{center}
\vspace{0.1mm}
\begin{center}
{\bf Institute of Nuclear Physics, Uzbekistan Academy of Sciences,
\\Ulugbek, Tashkent 100214, Uzbekistan}
\end{center}
\vspace{0.1mm}

\begin{center}
{\bf Abstract}
\end{center}

Spontaneous mirror symmetry violation is carried out in nature as the transition between the usual left (right)-handed and the mirror right (left)-handed spaces, in each of which the usual and mirror particles have the different lifetimes. As a consequence, all equations of motion in a unified field theory of elementary particles include the mass, energy and momentum as the matrices expressing the ideas of the left- and right-handed neutrinos are of long- and short-lived objects, respectively. These ideas require in principle to go away from the chiral definitions of the structure of matter fields taking into account that the Dirac matrices are, in the Weyl presentation, reduced to the matrices indicating to the absence in nature of a place for parity conservation but not allowing 
to follow the dynamical origination of its spontaneous violation. We discuss a theory in which a set comes forward at the new level, namely, at the level of set-matrix duality principle as a criterion for matrices. This connection gives the exact mathematical definitions of internally disclosed and undisclosed matrices, allowing to formulate three more definitions, three lemmas and two pairs of axioms. Thereby, it involves that there is no single matrix, for which an absolutely empty matrix would not exist. The sets of matrix elements and the matrices of set elements thus found unite 
all of matrix and set operations necessary for deciding the problems in a unified whole.  

\vspace{0.3cm}

\vspace{0.3cm}

\vspace{0.8cm}
\noindent
{\bf 1. Introduction}
\vspace{0.4cm}

One structural set of the innate properties of matter, which was not internally disclosed before 
the creation of the first-initial unified field theory, is spontaneous mirror symmetry violation. 
It is not surprising therefore that in the form as it was accepted, neither of the quantum mechanical equations depending on the mass, energy, and momentum is in a state to describe 
the elementary objects by the mirror symmetry laws.

At the same time, nature itself relates the same left or right spin state of a particle even, in the case of the neutrino $(\nu_{l}=\nu_{e},$ $\nu_{\mu},$ $\nu_{\tau}, ...),$ to corresponding component of its antiparticle. It constitutes [1] herewith an individual paraneutrino 
$$(\nu_{lL}, {\bar \nu_{lR}}), \, \, \, \,
(\nu_{lR}, {\bar \nu_{lL}}),$$
confirming the availability in it of the transitions between the left and the right.

However, as was accepted in the standard electroweak model [2-4], their existence contradicts one of its postulates that in nature the right-handed neutrinos are absent. Instead it includes the right components of leptons $(l=e,$ $\mu,$ $\tau, ...)$ as the usual singlets. 

But if we take into account that the mass, energy, and momentum of any of elementary particles unite all symmetry laws in a unified whole, then to any type of lepton corresponds in their spectra [5] a kind of neutrino [6]. Thereby, they describe a situation when mirror symmetry violation spontaneously originates in any [1,7] of interconversions 
\begin{equation}
l_{L}\leftrightarrow l_{R}, \, \, \, \, \overline{l}_{R}\leftrightarrow \overline{l}_{L},
\label{1}
\end{equation}
\begin{equation}
\nu_{lL}\leftrightarrow \nu_{lR}, \, \, \, \, {\bar \nu}_{lR}\leftrightarrow {\bar \nu}_{lL}
\label{2}
\end{equation} 
by the same mechanism. Such a mechanism can, for example, be simultaneous change of the mass, 
energy, and momentum of a particle at its transition from one spin state into another. It reflects 
the availability of the usual left (right)-handed and the mirror right (left)-handed Minkowski 
space-times. Therefore, to understand the nature of elementary particles at the new dynamical level, one must use each interconversion of (\ref{1}) and (\ref{2}) as the transition between the usual 
and the mirror spaces [8], in each of which the usual and mirror particles have the different masses, energies, and momenta. This connection expresses, in the case of the C-invariant Dirac neutrino, the idea about that the left-handed neutrino and the right-handed antineutrino are of long-lived leptons of C-invariance, and the right-handed neutrino and the left-handed antineutrino refer to short-lived C-even fermions.

The unidenticality of lifetimes $\tau_{s}$ and space-time coordinates $(t_{s}, {\bf x}_{s})$ of left $(s=L=-1)$ and right $(s=R=+1)$ types of elementary objects of C-parity establishes in addition the full spin structure of all equations of motion in a unified field theory of particles with a nonzero spin in which the mass, energy, and momentum are predicted as the matrices
\begin{equation}
m_{s}={{m_{V} \, \, \, \, 0}\choose{\ 0 \, \, \, \, \ m_{V}}}, \, \, \, \,
E_{s}={{E_{V} \, \, \, \, 0}\choose{\ 0 \, \, \, \, \ E_{V}}}, \, \, \, \,
{\bf p}_{s}={{{\bf p}_{V} \, \, \, \, 0}\choose{\ 0 \, \, \, \, \ {\bf p}_{V}}},
\label{3}
\end{equation}
\begin{equation}
m_{V}={{m_{L} \, \, \, \, 0}\choose{\ 0 \, \, \, \, \ m_{R}}}, \, \, \, \,
E_{V}={{E_{L} \, \, \, \, 0}\choose{\ 0 \, \, \, \, \ E_{R}}}, \, \, \, \,
{\bf p}_{V}={{{\bf p}_{L} \, \, \, \, 0}\choose{\ 0 \, \, \, \, \ {\bf p}_{R}}}.
\label{4}
\end{equation}

Such a presentation of $m_{s},$ $E_{s},$ and ${\bf p}_{s}$ is of course intimately connected with the character of their compound structure depending on a vector $(V)$ nature [8] of the same space-time, where there exist C-invariant particles.

However, among the sets of C-even objects there are no C-odd particles. Their mass, energy, and momentum do not coincide with (\ref{3}) and (\ref{4}), since in them appears an axial-vector $(A)$ nature [9] of the same space-time, where there exist C-noninvariant particles. They can therefore 
be expressed in the form
\begin{equation}
m_{s}={{0 \ \, \, \, \, m_{A}}\choose{m_{A} \, \, \, \, \ 0}}, \, \, \, \,
E_{s}={{0 \ \, \, \, \, E_{A}}\choose{E_{A} \, \, \, \, \ 0}}, \, \, \, \,
{\bf p}_{s}={{0 \ \, \, \, \, {\bf p}_{A}}\choose{{\bf p}_{A} \, \, \, \, \ 0}},
\label{5}
\end{equation}
\begin{equation}
m_{A}={{m_{L} \, \, \, \, 0}\choose{\ 0 \, \, \, \, \ m_{R}}}, \, \, \, \,
E_{A}={{E_{L} \, \, \, \, 0}\choose{\ 0 \, \, \, \, \ E_{R}}}, \, \, \, \,
{\bf p}_{A}={{{\bf p}_{L} \, \, \, \, 0}\choose{\ 0 \, \, \, \, \ {\bf p}_{R}}}.
\label{6}
\end{equation}

This difference corresponds in nature to a separation [10] of elementary currents with respect 
to C-operation, because it admits the existence of C-even and C-odd types of particles of vector 
$(V)$ and axial-vector $(A)$ masses, energies, and momenta.

It is also relevant to use [8,9] their sizes as the quantum operators 
\begin{equation}
m_{s}=-i\frac{\partial}{\partial \tau_{s}}, \, \, \, \,  
E_{s}=i\frac{\partial}{\partial t_{s}}, \, \, \, \,  
{\bf p}_{s}=-i\frac{\partial}{\partial {\bf x}_{s}}.
\label{7}
\end{equation}

Furthermore, if the investigated and the used objects are simultaneously both C-even and C-odd neutrinos, a motion of all types of particles with the spin $1/2$ and the four-component wave function $\psi_{s}(t_{s}, {\bf x}_{s})$ may in a mirror world [8,9] be described by a latent 
united equation 
\begin{equation}
i\frac{\partial}{\partial t_{s}}\psi_{s}=\hat H_{s}\psi_{s},
\label{8}  
\end{equation} 
which states that 
\begin{equation}
\hat H_{s}=\alpha \cdot\hat {\bf p_{s}}+\beta m_{s}.
\label{9}
\end{equation}

Using a unity $I$ matrix, the Pauli spin $\sigma$ matrices 
\begin{equation}
\sigma_{x}={{0\ \, \, \, \, \ 1}\choose{1\ \, \, \, \, \ 0}}, \, \, \, \,
\sigma_{y}={{0\ -i}\choose{i\ \, \, \, \, \ 0}}, \, \, \, \,
\sigma_{z}={{1\ \, \, \, \, \ 0}\choose{0\ -1}}
\label{10}
\end{equation}
and taking into account the standard presentation of the Dirac [10], for $\alpha,$ $\beta,$ and 
$\gamma_{5},$ we have
\begin{equation}
\alpha= {{0 \ \, \, \, \, \sigma}\choose{\sigma \, \, \, \, \ 0}}, \, \, \, \,
\beta={{I \ \, \, \, \, \ 0}\choose{0 \ -I}}, \, \, \, \,
\gamma_{5}={{0 \, \, \, \, \ I}\choose{I \ \, \, \, \, \ 0}}.
\label{11}
\end{equation}

At a choice of the above matrices, the solutions of an equation (\ref{8}) reflect, in the case of both vector [8] and axial-vector [9] types of fermions, the same characteristic features of quantum mechanical helicity operator $\sigma{\bf p}_{s}=s|{\bf p}_{s}|,$ which indicate to a unified principle that 
\begin{equation}
\sigma{\bf p_{L}}=-|{\bf p}_{L}|, \, \, \, \, \sigma{\bf p_{R}}=|{\bf p}_{R}|.
\label{12}
\end{equation}

However, the fact that the classical Dirac equation includes the usual mass, energy, and 
momentum would seem to require one to raise the question about the structure of matrices 
$\alpha=\gamma_{5}\sigma$ and $\beta$ as to whether there exists any connection between their definition and the spontaneous mirror symmetry violation. At the same time, a very form of each 
of matrices (\ref{11}) is not singular from the mathematical viewpoint.

Therefore, it seems that $\alpha,$ $\beta,$ and $\gamma_{5}$ can in the chiral presentation of 
the Weyl [11] have the following form:
\begin{equation}
\alpha= {{\sigma \ \, \, \, \, \ 0}\choose{0 \, \, \, \, -\sigma}}, \, \, \, \,
\beta={{0 \, \, \, \, \ I}\choose{I \ \, \, \, \, \ 0}}, \, \, \, \,
\gamma_{5}={{I \ \, \, \, \, \ 0}\choose{0 \ -I}}.
\label{13}
\end{equation}

Our purpose in a given work is to follow the mathematical logic of so far unobserved relations between the solutions of an equation (\ref{8}) in the presence in it of either the matrices (\ref{11}) or the matrices (\ref{13}) both from the point of view of vector C-invariant types 
of neutrinos and on the basis of C-noninvariance of axial-vector types of neutrinos. They will 
be illuminated in sections 3, 4, and 5 including their generality. This does not exclude of course from the discussion a theory in which a set comes forward as a criterion for matrices, and a matrix is none other than a criterion for sets. It will be presented in sections 6, 7, and 8, recognizing that in them, a role of hitherto latent set-matrix duality principle appears. In section 9, we make some concluding remarks.

However, we cannot define their structure until the very logical set of the used mathematical concepts has an exact definition.

\vspace{0.8cm}
\noindent
{\bf 2. Preliminaries}
\\{\bf 2.1 \it Internally disclosed sets of an imaginary space}
\vspace{0.4cm}

To define at the new level the mathematical notion of a set, one must refer to the mathematical notion of what unites [12] all of its elements in a unified whole as a latent object of unification, because it without this refers to fully casual sets in which there is no internal disclosure. A class with a latent object of unification of its elements refers to fully regular sets having internal disclosure. In any of them [13,14], an empty class [15], not having any element is necessarily present.

{\bf Definition 1.} A latent algebraical object of unification of one set is the second set, such 
that it consists of conserving sizes of the same defined symmetry of elements of both sets.

{\bf Definition 2.} A latent geometrical object of unification of one set is the second set, such that it consists of conserving points of the same defined line of elements of both sets.

{\bf Definition 3.} The sets are called fully regular or internally disclosed ones if each of 
them corresponds to one pair of the algebraical and geometrical objects of latent unification 
of its elements.

{\bf Definition 4.} The sets are called fully casual or internally undisclosed ones if none of them has neither the algebraical nor geometrical objects of latent unification of its elements. 

{\bf Lemma 1 (Theorem on the smallest nonempty set).} No single fully regular set from the same element exists without an internally disclosed set of a higher cardinality.

Another of the structural contradictions between these sets [13,14] is that a space comes 
forward as a set of those objects, the existence of which follows from the fundamental 
mathematical principles.

But there is no single nonempty space in nature, for which a definitely symmetrical number 
axis [13,14] would not correspond. If such a space is an imaginary one, the latter makes it possible to introduce the notion of an imaginary number axis [13], allowing one to formulate and prove the theorem on the basis of its internal disclosure. 

To show this, we must at first define [13] the mathematical notion of an imaginary space expressing the ideas of the symmetry laws about how to each type of positive (negative) number on an imaginary axis corresponds a kind of higher (lower) point.

{\bf Definition 5.} A nonempty space is called an imaginary one if and only if it consists of 
the higher and lower points of infinitely many selected systems of imaginary axes with a general center of symmetry.

{\bf Definition 6.} A number axis is called an imaginary one if it has higher and lower 
imaginary points relative to its center of symmetry.

{\bf Theorem 1.} To each pair of higher and lower imaginary points corresponds one pair of objects such that together they constitute a system of two sets from two elements of an imaginarily 
defined symmetry.

{\bf Proof of Theorem 1.} There is no single point in an imaginary number axis
\begin{equation}
..., -(n+1)i, \, \, \, \, -ni, ..., -3i, \, \, \, \, -2i, \, \, \, \, -i, \, \, \, \, 
0, \, \, \, \, i, \, \, \, \, 2i, \, \, \, \, 3i, ..., ni, \, \, \, \, (n+1)i, ...,
\label{14}
\end{equation}
for which, according to Theorem 1, a kind of imaginary object would not correspond. It characterizes herewith any higher point from 
\begin{equation}
i, \, \, \, \, 2i, \, \, \, \, 3i, ..., ni, \, \, \, \, (n+1)i, ...
\label{15}
\end{equation}
by one his imaginary object from
\begin{equation}
c_{1\Delta}, \, \, \, \, c_{2\Lambda}, \, \, \, \, c_{3\Pi}, ..., 
c_{n\Upsilon}, \, \, \, \, c_{(n+1)\Phi}, ....
\label{16}
\end{equation}

Each of indices of distinction $P=\Delta,$ $\Lambda,$ $\Pi, ...,$ $\Upsilon,$ $\Phi, ...$ corresponds in an imaginary space to one of the existing types of symmetries. 

Exactly the same one can describe any lower point from
\begin{equation}
-i, \, \, \, \, -2i, \, \, \, \, -3i, ..., -ni, \, \, \, \, -(n+1)i, ...
\label{17}
\end{equation}
by its corresponding imaginary object from
\begin{equation}
d_{1\Delta}, \, \, \, \, d_{2\Lambda}, \, \, \, \, d_{3\Pi}, ..., 
d_{n\Upsilon}, \, \, \, \, d_{(n+1)\Phi}, ...
\label{18}
\end{equation}
such that each index $P$ is responsible for distinction of one pair of imaginary objects from 
(\ref{16}) and (\ref{18}) among all others.

If now group (\ref{15}) and (\ref{17}), the sequence
\begin{equation}
i, \, \, \, \, -i, \, \, \, \, 2i, \, \, \, \, -2i, \, \, \, \, 3i, \, \, \, \, -3i, ..., 
ni, \, \, \, \, -ni, \, \, \, \, (n+1)i, \, \, \, \, -(n+1)i, ...
\label{19}
\end{equation}
requires one to group (\ref{16}), (\ref{18}) and constitute such a sequence as
\begin{equation}
c_{1\Delta}, \, \, \, \, d_{1\Delta}, \, \, \, \, 
c_{2\Lambda}, \, \, \, \, d_{2\Lambda}, \, \, \, \, c_{3\Pi}, \, \, \, \, d_{3\Pi}, ...,  
c_{n\Upsilon}, \, \, \, \, d_{n\Upsilon}, \, \, \, \, c_{(n+1)\Phi}, \, \, \, \, d_{(n+1)\Phi}, ....
\label{20}
\end{equation}

Thus, to each of sets of imaginary points
\begin{equation}
\{i, \, \, \, \, -i\}, \, \, \, \, \{2i, \, \, \, \, -2i\}, \, \, \, \, 
\{3i, \, \, \, \, -3i\}, ..., \{ni, \, \, \, \, -ni\}, \, \, \, \, 
\{(n+1)i, \, \, \, \, -(n+1)i\}, ...,
\label{21}
\end{equation}
as stated in Theorem 1, corresponds one and only one from sets of imaginary objects
\begin{equation}
\{c_{1\Delta}, \, \, \, \, d_{1\Delta}\}, \, \, \, \, 
\{c_{2\Lambda}, \, \, \, \, d_{2\Lambda}\}, \, \, \, \, \{c_{3\Pi}, \, \, \, \, d_{3\Pi}\}, ...,  
\{c_{n\Upsilon}, \, \, \, \, d_{n\Upsilon}\}, \, \, \, \, 
\{c_{(n+1)\Phi}, \, \, \, \, d_{(n+1)\Phi}\}, ....
\label{22}
\end{equation} 

Furthermore, if one-to-one correspondence between the series (\ref{21}) and
\begin{equation}
\{i, \, \, \, \, -i\}, \, \, \, \, \{i, \, \, \, \, -i\}, \, \, \, \, 
\{i, \, \, \, \, -i\}, ..., \{i, \, \, \, \, -i\}, \, \, \, \, \{i, \, \, \, \, -i\}, ...
\label{23}
\end{equation}
is of the fundamental mathematical principles, $i$ and $-i$ undoubtedly refer only to the 
individual numbers of each conserving size 
${\EuScript P}=\varDelta,$ $\varLambda,$ $\varPi, ..., $ $\varUpsilon,$ $\varPhi, ...$ 
of existing types of symmetries $P$ of a very imaginary space.

On this basis, (\ref{22}) and (\ref{23}) constitute a united system of sets
\begin{equation}
\left\{
\begin{array}{lll}
F=\{c_{1\Delta}, \, \, \, \, d_{1\Delta}\},&&\\
\varDelta=\{i, \, \, \, \, -i\},&&\\
\end{array}
\right.
\label{24}
\end{equation}
\begin{equation}
\left\{
\begin{array}{lll}
V=\{c_{n\Upsilon}, \, \, \, \, d_{n\Upsilon}\},&&\\
\varUpsilon=\{i, \, \, \, \, -i\},&&\\
\end{array}\right.
\label{25}
\end{equation}
\begin{equation}
\left\{
\begin{array}{lll}
W=\{c_{(n+1)\Phi}, \, \, \, \, d_{(n+1)\Phi}\},&&\\
\varPhi=\{i, \, \, \, \, -i\}.&&\\
\end{array}\right.
\label{26}
\end{equation}

The second set in each of systems (\ref{24})-(\ref{26}) is none other than a class, the elements of which, as was mentioned in Definition 1, satisfy one of 
\begin{equation}
\sum\varDelta=const, ..., \sum\varUpsilon=const, \, \, \, \, \sum\varPhi=const, ....
\label{27}
\end{equation}

We establish in an imaginary space a general system of sets 
\begin{equation}
\left\{
\begin{array}{lll}
I=\{F, \, \, \, \,O, \, \, \, \, Q, ..., V, \, \, \, \, W, ...\},&&\\
{\EuScript P}=\{\varDelta, \, \, \, \, \varLambda, \, \, \, \, 
\varPi, ..., \varUpsilon, \, \, \, \, \varPhi, ...\}&&\\
\end{array}\right.
\label{28}
\end{equation}
having an internal disclosure 
\begin{equation}
\sum {\EuScript P}=const.
\label{29}
\end{equation}

As seen, the very unity of all types of symmetry laws of an imaginary space comes forward in a system (\ref{28}) as the same defined symmetry $P$ with the same conserving size ${\EuScript P},$ thus confirming the validity of Theorem 1 and all of its implications. ${\square}$  

Theorem 1 characterizes at
$$NP=1\Delta, \, \, \, \, 2\Lambda, \, \, \, \, 3\Pi, ..., n\Upsilon, \, \, \, \, 
(n+1)\Phi, ...$$
each element from $c_{NP}$ and $d_{NP}$ or subclass from $\{c_{NP}\}$ and $\{d_{NP}\}$ of a set $I$ of a general system (\ref{28}), respectively, by an individual imaginary number
\begin{equation}
{\EuScript P}=\left\{
{\begin{array}{l}
{+i\quad \mbox{for}\quad c_{NP},}\\
{-i\quad \mbox{for}\quad d_{NP},}\\
{\, \, \, \nexists\quad \mbox{for}\quad \mbox{remaining objects}}\\
\end{array}}\right.
\label{30}
\end{equation}
or, according to Lemma 1, by an individual subset 
\begin{equation}
{\EuScript P}=\left\{
{\begin{array}{l}
{\{+i\}\quad \mbox{for}\quad \{c_{NP}\},}\\
{\{-i\}\quad \mbox{for}\quad \{d_{NP}\},}\\
{\, \, \, \, \varnothing\quad \, \, \, \, \mbox{for}\quad \mbox{remaining subclasses}}\\
\end{array}}\right.
\label{31}
\end{equation}
such that a constancy law of the sum (\ref{29}), which states its algebraical disclosure is never violated. Any of the Definitions 3 and 4 together with the Definition 1 says herewith in favor 
of a kind of Theorem.

{\bf Lemma 2 (Theorem on an internal disclosure algebraical logic).} There is no algebraical disclosure in a set without a strictly defined symmetry of elements.

{\bf Lemma 3 (Theorem on an internal undisclosure algebraical logic).} There is no algebraical undisclosure in a set without a strictly defined antisymmetry of elements.

Furthermore, if it turns out that a constancy law of the sum (\ref{29}) confirms simultaneously
the validity of both Lemmas 1 and 2, the legality of Lemma 3 follows from its violation 
\begin{equation}
\sum {\EuScript P}\neq const.
\label{32}
\end{equation}

Each of the Definitions 3 and 4 jointly with the Definition 2 indicates herewith on the existence 
of a kind of pair of Theorems.

{\bf Lemma 4 (Theorem on an internal disclosure geometrical logic).} There is no geometrical disclosure in a set without a definitely symmetrical line of elements.

{\bf Lemma 5 (Theorem on an internal undisclosure geometrical logic).} There is no geometrical undisclosure in a set without a definitely antisymmetrical line of elements.

Insofar as the proof of Lemmas 4 and 5 is concerned, here we must keep in mind [14] that the 
very imaginary space characterizes each commutative pair of each imaginary number axis both 
by the imaginative curved and straight lines.

{\bf Definition 7.} A line is called an imaginative curved one if it unites all points with images of one and only one of objects of each commutative pair. 

{\bf Definition 8.} A line is called an imaginative straight one if it unites selected
points with images of each of objects of each commutative pair. 

{\bf Lemma 6 (Theorem on a commutative pair mathematical logic).} There is no mathematical 
disclosure in a set without commutative pairs of elements.

{\bf Lemma 7 (Theorem on an anticommutative pair mathematical logic).} There is no mathematical undisclosure in a set without anticommutative pairs of elements.

The validity of both Lemmas 6 and 7 follows from the fact that the Definitions 3 and 4 together with the Definitions 1 and 2 unite each of Lemmas 2 and 3 with the corresponding from the Lemmas 4 and 5, thus confirming the validity of one more pair of highly importanr Theorems.

{\bf Lemma 8 (Theorem on an internal disclosure mathematical logic).} There is no mathematical disclosure in a set without a definitely symmetrical line of elements of strictly defined symmetry.

{\bf Lemma 9 (Theorem on an internal undisclosure mathematical logic).} There is no mathematical undisclosure in a set without a definitely antisymmetrical line of elements of strictly defined antisymmetry.

Definitions 3 and 4 jointly with the Lemmas 8 and 9 express, for each of the internally disclosed and undisclosed types of sets, the idea of a kind of Acsioms.

{\bf Axiom 1.} An internal disclosure of a set is none other than its
mathematical disclosure.
 
{\bf Axiom 2.} An internal undisclosure of a set is none other than its 
mathematical undisclosure.

\vspace{0.8cm} 
\noindent
{\bf 2.2 \it Fully regular sets of a real space}
\vspace{0.4cm}

Lemmas 1-9, Definitions 1-8, and Acsioms 1 and 2 have the generality [13] such that requires one 
to follow the logic of a set consisting of objects of a real space [14]. But there is no imaginary 
number axis in a real space. Instead it defines a real number axis [16,17] at the new level, namely, at the level of real space, allowing one to formulate and prove the theorem on the basis of its internal disclosure.

To investigate further, we must at first define [14] the mathematical notion of a real space expressing the ideas of the symmetry laws about how to each type of positive (negative) number 
on a real axis corresponds a kind of right (left) point.

{\bf Definition 9.} A nonempty space is called a real one if and only if it consists of the right and left points of infinitely many selected systems of real axes with a general center of symmetry. 

{\bf Definition 10.} A number axis is called a real one if it has right and left real points 
relative to its center of symmetry. 

{\bf Theorem 2.} To each pair of right and left real points corresponds one pair of objects 
such that together they constitute a system of two sets from two elements of a really defined symmetry. 

{\bf Proof of Theorem 2.} Each point of a real number axis
\begin{equation}
..., -(n+1), \, \, \, \, -n, ..., -3, \, \, \, \, -2, \, \, \, \, -1, \, \, \, \, 
0, \, \, \, \, 1, \, \, \, \, 2, \, \, \, \, 3, ..., n, \, \, \, \, (n+1), ...,
\label{33}
\end{equation}
as stated in Theorem 2, must have his own real object. It describes herewith any right 
point from
\begin{equation}
1, \, \, \, \, 2, \, \, \, \, 3, ..., n, \, \, \, \, (n+1), ...
\label{34}
\end{equation}
by one corresponding real object from 
\begin{equation}
a_{1B}, \, \, \, \, a_{2L}, \, \, \, \, a_{3X}, ..., a_{nY}, \, \, \, \, a_{(n+1)Z}, ....
\label{35}
\end{equation}

One index of a distinction from $E=B,$ $L,$ $X, ..., $ $Y,$ $Z, ...$ denotes in a real space
one of all types of its symmetries. 

In a similar way one can characterize each left point from
\begin{equation}
-1, \, \, \, \, -2, \, \, \, \, -3, ..., -n, \, \, \, \, -(n+1), ...
\label{36}
\end{equation}
by its own real object from 
\begin{equation}
b_{1B}, \, \, \, \, b_{2L}, \, \, \, \, b_{3X}, ..., b_{nY}, \, \, \, \, b_{(n+1)Z}, ...
\label{37}
\end{equation}
such that any index $E$ distinguishes one pair of real objects of (\ref{35}) and (\ref{37})
from all the remaining ones.

One of the admissible groupings of (\ref{34}) and (\ref{36}) gives the sequence
\begin{equation}
1, \, \, \, \, -1, \, \, \, \, 2, \, \, \, \, -2, \, \, \, \, 3, \, \, \, \, -3, ..., 
n, \, \, \, \, -n, \, \, \, \, n+1, \, \, \, \, -(n+1), ....
\label{38}
\end{equation}

In the same way, one can group (\ref{35}), (\ref{37}) and constitute the sequence
\begin{equation}
a_{1B}, \, \, \, \, b_{1B}, \, \, \, \, a_{2L}, \, \, \, \, b_{2L}, \, \, \, \, 
a_{3X}, \, \, \, \, b_{3X}, ..., a_{nY}, \, \, \, \, b_{nY}, \, \, \, \, 
a_{(n+1)Z}, \, \, \, \, b_{(n+1)Z}, ....
\label{39}
\end{equation}

They in favor of that to each of sets of real points 
\begin{equation}
\{1, \, \, \, \, -1\}, \, \, \, \, \{2, \, \, \, \, -2\}, \, \, \, \, 
\{3, \, \, \, \, -3\}, ..., \{n, \, \, \, \, -n\}, \, \, \, \, 
\{(n+1), \, \, \, \, -(n+1)\}, ...,
\label{40}
\end{equation}
according to Theorem 2, corresponds one set of real objects 
\begin{equation}
\{a_{1B}, \, \, \, \, b_{1B}\}, \, \, \, \, \{a_{2L}, \, \, \, \, b_{2L}\}, \, \, \, \, 
\{a_{3X}, \, \, \, \, b_{3X}\}, ..., \{a_{nY}, \, \, \, \, b_{nY}\}, \, \, \, \, 
\{a_{(n+1)Z}, \, \, \, \, b_{(n+1)Z}\}, ....
\label{41}
\end{equation} 

There exists of course one-to-one correspondence between the series (\ref{40}) and
\begin{equation}
\{1, \, \, \, \, -1\}, \, \, \, \, \{1, \, \, \, \, -1\}, \, \, \, \, 
\{1, \, \, \, \, -1\}, ..., \{1, \, \, \, \, -1\}, \, \, \, \, 
\{1, \, \, \, \, -1\}, ...
\label{42}
\end{equation}
in which appears a part of the fundamental mathematical principles and that, consequently, 
$1$ and $-1$ are of the individual numbers of each conserving size 
${\EuScript E}={\EuScript B},$ ${\EuScript L},$ ${\EuScript X}, ...,$ 
of all types of symmetries $E$ of a very real space.

Uniting (\ref{41}) with (\ref{42}), we get the following systems of sets:
\begin{equation}
\left\{
\begin{array}{lll}
A=\{a_{1B}, \, \, \, \, b_{1B}\},&&\\
{\EuScript B}=\{1, \, \, \, \, -1\},&&\\
\end{array}
\right.
\label{43}
\end{equation}
\begin{equation}
\left\{
\begin{array}{lll}
K=\{a_{nY}, \, \, \, \, b_{nY}\},&&\\
{\EuScript Y}=\{1, \, \, \, \, -1\},&&\\
\end{array}\right.
\label{44}
\end{equation}
\begin{equation}
\left\{
\begin{array}{lll}
T=\{a_{(n+1)Z}, \, \, \, \, b_{(n+1)Z}\},&&\\
{\EuScript Z}=\{1, \, \, \, \, -1\}.&&\\
\end{array}\right.
\label{45}
\end{equation}

The second class in any system of (\ref{43})-(\ref{45}) is none other than a set, which was mentioned in Definition 1, and therefore, its elements satisfy one of  
\begin{equation}
\sum {\EuScript B}=const, ..., \sum {\EuScript Y}=const, \, \, \, \, \sum {\EuScript Z}=const, ....
\label{46}
\end{equation}

We establish in a real space a general system of sets 
\begin{equation}
\left\{
\begin{array}{lll}
R=\{A, \, \, \, \, C, \, \, \, \, G, ..., K, \, \, \, \, T, ...\},&&\\
{\EuScript E}=\{{\EuScript B}, \, \, \, \, {\EuScript L}, \, \, \, \, 
{\EuScript X}, ..., {\EuScript Y}, \, \, \, \, {\EuScript Z}, ...\}&&\\
\end{array}\right.
\label{47}
\end{equation}
having an internal disclosure 
\begin{equation}
\sum {\EuScript E}=const.
\label{48}
\end{equation}

As we see, the very unity of the existing types of symmetry laws of a real space expresses, 
for a system (\ref{47}), the idea of the same defined symmetry $E$ with the same conserving 
size ${\EuScript E},$ thereby confirming the validity of Theorem 2 and all of its 
consequences. ${\square}$  

Theorem 2 characterizes at
$$NE=1B, \, \, \, \, 2L, \, \, \, \, 3X, ..., nY, \, \, \, \, (n+1)Z, ...$$
any element from $a_{NE}$ and $b_{NE}$ or subset from $\{a_{NE}\}$ and $\{b_{NE}\}$ of a set $R$
of a general system (\ref{47}), respectively, by an individual real number
\begin{equation}
{\EuScript E}=\left\{
{\begin{array}{l}
{+1\quad \mbox{for}\quad a_{NE},}\\
{-1\quad \mbox{for}\quad b_{NE},}\\
{\, \, \, \, \nexists\quad \mbox{for}\quad \mbox{remaining objects}}\\
\end{array}}\right.
\label{49}
\end{equation}
or, according to Lemma 1, by an individual subclass 
\begin{equation}
{\EuScript E}=\left\{
{\begin{array}{l}
{\{+1\}\quad \mbox{for}\quad \{a_{NE}\},}\\
{\{-1\}\quad \mbox{for}\quad \{b_{NE}\},}\\
{\, \, \, \, \, \varnothing\quad \, \, \, \, \mbox{for}\quad \mbox{remaining subclasses}}\\
\end{array}}\right.
\label{50}
\end{equation}
such that a constancy law of the sum (\ref{48}) expressing its algebraical disclosure 
is never violated. 

The legality in a real space of Lemma 3 unlike the valdity of both Lemmas 1 and 2 lmplyid from 
(\ref{48}) states that 
\begin{equation}
\sum {\EuScript E}\neq const
\label{51}
\end{equation}
refers to thous sets in which there is no algebraical disclosure.

Insofar as the geometrical disclosure or undisclosure of a set of real objects is concerned, 
it will appear, as was mentioned in Lemmas 4 and 5, in the symmetrical or antisymmetrical 
line dependence of elements.

The definitely symmetrical (antisymmetrical) line and strictly defined symmetry (antisymmetry) of elements of a set come forward in Lemma 8 (in Lemma 9) as the mathematical disclosure (undisclosure) of a real space.

\vspace{0.8cm}
\noindent
{\bf 3. Helicity operator of a vector field}
\vspace{0.4cm}

A notion about chiral symmetry introduced by Weyl is based factually on the presentation (\ref{13}), according to which, the matrix $\gamma_{5}$ becomes chirality operator having the same self-values 
as the helicity operator. In this case, it is expected that the solutions of an equation (\ref{8}) including (\ref{3}) and (\ref{4}) correspond in the presentations (\ref{11}) and (\ref{13}) to the most diverse forms of the same regularity of a C-invariant nature of vector $(V)$ types of fields.

To express the idea more clearly, we use a free particle with 
\begin{equation}
\psi_{s}=u_{s}({\bf p}_{s}, \sigma)e^{-ip_{s} \cdot{\it x}_{s}}, \, \, \, \, E_{s}>0.
\label{52}
\end{equation}

One can define its four-component spinor $u_{s}$ in the form
\begin{equation}
u_{s}=u^{(r)}=\left[\chi^{(r)}\atop u_{a}^{(r)}\right]
\label{53}
\end{equation}
in which
\begin{equation}
\chi^{(1)}=\left(1\atop 0\right), \, \, \, \, \chi^{(2)}=\left(0\atop 1\right),
\label{54}
\end{equation}
and the presence of an index $a$ in one of $u^{(r)}$ and $u_{a}^{(r)}$ is responsible 
for their distinction.

So, it is seen that (\ref{52}) together with (\ref{3}) and (\ref{13}) separates (\ref{8}) into 
\begin{equation}
E_{V}\chi^{(r)}=(\sigma{\bf p}_{V})\chi^{(r)}+m_{V}u_{a}^{(r)},
\label{55}
\end{equation}
\begin{equation}
E_{V}u_{a}^{(r)}=-(\sigma{\bf p}_{V})u_{a}^{(r)}+m_{V}\chi^{(r)}.
\label{56}
\end{equation}

Solving a given system concerning $\chi^{(r)}$ and $u_{a}^{(r)},$ but having in view of (\ref{53}), 
it can also be verified that (\ref{4}) and (\ref{54}) lead us from
\begin{equation}
u^{(r)}=\sqrt{E_{V}+(\sigma{\bf p}_{V})}
\left[\chi^{(r)}\atop \frac{m_{V}}{E_{V}+(\sigma{\bf p}_{V})}\chi^{(r)}\right]
\label{57}
\end{equation}
to their explicit form
\begin{equation}
u^{(1)}=\sqrt{E_{L}+(\sigma{\bf p}_{L})}
\left[
\begin{array}{c}
1\\ 0\\ \frac{m_{L}}{E_{L}+(\sigma{\bf p}_{L})}\\ 0
\end{array}
\right],
\label{58}
\end{equation}
\begin{equation}
u^{(2)}=\sqrt{E_{R}+(\sigma{\bf p}_{R})}
\left[
\begin{array}{c}
0\\ 1\\ 0\\ \frac{m_{R}}{E_{R}+(\sigma{\bf p}_{R})}
\end{array}
\right].
\label{59}
\end{equation}

At the same choice of a free particle and its four-component wave function, the solutions of an equation (\ref{8}) depending on (\ref{3}) and (\ref{4}) have [8] in the standard presentation (\ref{11}) the following structure:
\begin{equation}
u^{(1)}=\sqrt{E_{L}+m_{L}}
\left[
\begin{array}{c}
1\\ 0\\ \frac{(\sigma{\bf p}_{L})}{E_{L}+m_{L}}\\ 0
\end{array}
\right],
\label{60}
\end{equation}
\begin{equation}
u^{(2)}=\sqrt{E_{R}+m_{R}}
\left[
\begin{array}{c}
0\\ 1\\ 0\\ \frac{(\sigma{\bf p}_{R})}{E_{R}+m_{R}}
\end{array}
\right].
\label{61}
\end{equation}

From their point of view, the chiral presentation (\ref{13}) leading to (\ref{58}) and (\ref{59})
replaces the mass of a C-invariant particle by the operator of its helicity and vice versa. In other words, it requires one to make the replacements
\begin{equation}
m_{L,R}\rightarrow \sigma{\bf p_{L,R}}, \, \, \, \, \sigma{\bf p_{L,R}}\rightarrow m_{L,R}.
\label{62}
\end{equation}

In the same way one can solve the equation (\ref{8}) for the free antiparticle with
\begin{equation}
\psi_{s}=\nu_{s}({\bf p}_{s}, \sigma)e^{-ip_{s} \cdot{\it x}_{s}}, \, \, \, \, E_{s}<0.
\label{63}
\end{equation}

Its four-component spinor $\nu_{s}$ must have the form
\begin{equation}
\nu_{s}=\nu^{(r)}=\left[\nu_{a}^{(r)}\atop \chi^{(r)}\right].
\label{64}
\end{equation}

The availability of an index $a$ in one of $\nu^{(r)}$ and $\nu_{a}^{(r)}$ implies their difference. We see in addition that jointly with (\ref{3}) and (\ref{13}), the four-component wave function (\ref{63}) constitutes from (\ref{8}) a system of the two other equations
\begin{equation}
|E_{V}|\nu_{a}^{(r)}=-(\sigma{\bf p}_{V})\nu_{a}^{(r)}-m_{V}\chi^{(r)},
\label{65}
\end{equation}
\begin{equation}
|E_{V}|\chi^{(r)}=(\sigma{\bf p}_{V})\chi^{(r)}-m_{V}\nu_{a}^{(r)}.
\label{66}
\end{equation}

Inserting the second of its solutions 
\begin{equation}
\chi^{(r)}=\frac{-m_{V}}{|E_{V}|-(\sigma{\bf p}_{V})}\nu_{a}^{(r)}, \, \, \, \,
\nu_{a}^{(r)}=\frac{-m_{V}}{|E_{V}|+(\sigma{\bf p}_{V})}\chi^{(r)}.
\label{67}
\end{equation}
in (\ref{64}) and uniting the finding equality with (\ref{4}) and (\ref{54}), it is not difficult 
to show that
\begin{equation}
\nu^{(1)}=\sqrt{|E_{L}|+(\sigma{\bf p}_{L})}
\left[
\begin{array}{c}
\frac{-m_{L}}{|E_{L}|+(\sigma{\bf p}_{L})}\\ 0\\ 1\\ 0
\end{array}
\right],
\label{68}
\end{equation}
\begin{equation}
\nu^{(2)}=\sqrt{|E_{R}|+(\sigma{\bf p}_{R})}
\left[
\begin{array}{c}
0\\ \frac{-m_{R}}{|E_{R}|+(\sigma{\bf p}_{R})}\\ 0\\ 1
\end{array}
\right].
\label{69}
\end{equation}

If choose the standard presentation (\ref{11}), at which the matrix $\gamma_{5}$ is not chirality operator, then for the same case of a free antiparticle when (\ref{3}), (\ref{4}), and (\ref{53}) 
refer to it, one can establish the compound structure of both types of solutions of an equation (\ref{8}) in the disclosed form [8] by the following manner:
\begin{equation}
\nu^{(1)}=\sqrt{|E_{L}|+m_{L}}
\left[
\begin{array}{c}
\frac{-(\sigma{\bf p}_{L})}{|E_{L}|+m_{L}}\\ 0\\ 1\\ 0
\end{array}
\right],
\label{70}
\end{equation}
\begin{equation}
\nu^{(2)}=\sqrt{|E_{R}|+m_{R}}
\left[
\begin{array}{c}
0\\ \frac{-(\sigma{\bf p}_{R})}{|E_{R}|+m_{R}}\\ 0\\ 1
\end{array}
\right].
\label{71}
\end{equation}

Their comparison with (\ref{68}) and (\ref{69}) convinces us in the validity of (\ref{62}) once more, confirming that the chiral presentation (\ref{13}) replaces the helicity operator of a C-invariant antiparticle by its mass and vice versa. 

\vspace{0.8cm}
\noindent
{\bf 4. Helicity operator of an axial-vector field}
\vspace{0.4cm}

Between the vector and the axial-vector spaces [18] there exists a range of fundamental differences, which require the unification of elementary particles with respect to C-operation. However, nature, by itself, does not separate [8,9] each of these forms of Minkowski spaces into left and right spaces, and the transitions between the different spin states are carried out in it spontaneously 
by a mirror symmetry violation. It chooses herewith the mass, energy, and momentum matrices so that to the case of C-even [19] or C-odd [20] types of particles corresponds in their unified field theory a kind of equation of motion.

Therefore, from its point of view, it should be expected that an equation (\ref{8}) including (\ref{5}) and (\ref{6}) describe in the presentations (\ref{11}) and (\ref{13}) the most diverse 
forms of the same regularity of a C-noninvariant nature of axial-vector $(A)$ types of fields.

To elucidate these ideas, we use (\ref{52})-(\ref{54}) for the free particles of C-oddity. Then it 
is possible, for example, (\ref{52}) in the presence of (\ref{5}) and (\ref{13}) transforms (\ref{8}) into the system
\begin{equation}
E_{A}u_{a}^{(r)}=(\sigma{\bf p}_{A})u_{a}^{(r)}+m_{A}\chi^{(r)},
\label{72}
\end{equation}
\begin{equation}
E_{A}\chi^{(r)}=-(\sigma{\bf p}_{A})\chi^{(r)}+m_{A}u_{a}^{(r)}.
\label{73}
\end{equation}

It establishes the corresponding connections
\begin{equation}
u_{a}^{(r)}=\frac{m_{A}}{E_{A}-(\sigma{\bf p}_{A})}\chi^{(r)}, \, \, \, \,
\chi^{(r)}=\frac{m_{A}}{E_{A}+(\sigma{\bf p}_{A})}u_{a}^{(r)}.
\label{74}
\end{equation}

The first of them together with (\ref{54}) gives the right to define the four-component spinors $u^{(r)}$ for C-odd types of neutrinos
\begin{equation}
u^{(1)}=\sqrt{E_{L}-(\sigma{\bf p}_{L})}
\left[
\begin{array}{c}
1\\ 0\\ \frac{m_{L}}{E_{L}-(\sigma{\bf p}_{L})}\\ 0
\end{array}
\right],
\label{75}
\end{equation}
\begin{equation}
u^{(2)}=\sqrt{E_{R}-(\sigma{\bf p}_{R})}
\left[
\begin{array}{c}
0\\ 1\\ 0\\ \frac{m_{R}}{E_{R}-(\sigma{\bf p}_{R})}
\end{array}
\right].
\label{76}
\end{equation}

However, in the C-noninvariant case of a free particle, an equation (\ref{8}) depending on (\ref{5}) and (\ref{6}) can in the standard presentation (\ref{11}) have [9] the following solutions:
\begin{equation}
u^{(1)}=\sqrt{E_{L}-m_{L}}
\left[
\begin{array}{c}
1\\ 0\\ \frac{(\sigma{\bf p}_{L})}{E_{L}-m_{L}}\\ 0
\end{array}
\right],
\label{77}
\end{equation}
\begin{equation}
u^{(2)}=\sqrt{E_{R}-m_{R}}
\left[
\begin{array}{c}
0\\ 1\\ 0\\ \frac{(\sigma{\bf p}_{R})}{E_{R}-m_{R}}
\end{array}
\right].
\label{78}
\end{equation}

As we see, the chiral presentation (\ref{13}) establishing (\ref{75}) and (\ref{76}) replaces the mass of a C-noninvariant particle by the operator of its helicity and vice versa. 

Unification of (\ref{8}) with (\ref{5}) and (\ref{13}) at the discussion of a C-odd antiparticle described by (\ref{63}) suggests a system
\begin{equation}
|E_{A}|\chi^{(r)}=-(\sigma{\bf p}_{A})\chi^{(r)}-m_{A}\nu_{a}^{(r)},
\label{79}
\end{equation}
\begin{equation}
|E_{A}|\nu_{a}^{(r)}=(\sigma{\bf p}_{A})\nu_{a}^{(r)}-m_{A}\chi^{(r)}.
\label{80}
\end{equation}

Insertion of the first of its solutions $\nu_{a}^{(r)}$ and $\chi^{(r)}$ in (\ref{64}) allows 
one to conclude that
\begin{equation}
\nu^{(r)}=\sqrt{|E_{A}|-(\sigma{\bf p}_{A})}
\left[\frac{-m_{A}}{|E_{A}|-(\sigma{\bf p}_{A})}\chi^{(r)}\atop \chi^{(r)}\right].
\label{81}
\end{equation}

Because of (\ref{6}) and (\ref{54}), the latent structure of $\nu^{(r)}$ is disclosed in the 
following its sizes:
\begin{equation}
\nu^{(1)}=\sqrt{|E_{L}|-(\sigma{\bf p}_{L})}
\left[
\begin{array}{c}
\frac{-m_{L}}{|E_{L}|-(\sigma{\bf p}_{L})}\\ 0\\ 1\\ 0
\end{array}
\right], 
\label{82}
\end{equation}
\begin{equation}
\nu^{(2)}=\sqrt{|E_{R}|-(\sigma{\bf p}_{R})}
\left[
\begin{array}{c}
0\\ \frac{-m_{R}}{|E_{R}|-(\sigma{\bf p}_{R})}\\ 0\\ 1
\end{array}
\right].
\label{83}
\end{equation}

But in the standard presentation (\ref{11}), the equation (\ref{8}) for the same C-odd antiparticle with (\ref{5}), (\ref{6}), and (\ref{63}) establishes [9] the two other spinors 
\begin{equation}
\nu^{(1)}=\sqrt{|E_{L}|-m_{L}}
\left[
\begin{array}{c}
\frac{-(\sigma{\bf p}_{L})}{|E_{L}|-m_{L}}\\ 0\\ 1\\ 0
\end{array}
\right],
\label{84}
\end{equation}
\begin{equation}
\nu^{(2)}=\sqrt{|E_{R}|-m_{R}}
\left[
\begin{array}{c}
0\\ \frac{-(\sigma{\bf p}_{R})}{|E_{R}|-m_{R}}\\ 0\\ 1
\end{array}
\right].
\label{85}
\end{equation}

At the action of (\ref{62}) they coincide with the corresponding values from (\ref{82}), (\ref{83}) and that, consequently, the behavior of the chiral presentation (\ref{13}) is not 
changed even at a choice of a C-noninvariant antiparticle.

\vspace{0.8cm}
\noindent
{\bf 5. What helicity operators say about matrices
of a Dirac eqation}
\vspace{0.4cm}

Turning again to the structure and the component of the finding wave functions, we remark that 
the sign in front of a size of $m_{L,R}$ in $u_{a}^{(1)},$ $u_{a}^{(2)},$ $\nu_{a}^{(1)},$ and 
$\nu_{a}^{(2)}$ for C-even and C-odd particles does not coincide. This, however, does not exclude [8,9] the fact that $u^{(1)},$ $\chi^{(1)},$ and $u_{a}^{(1)}$ describe the left-handed neutrino, 
and $u^{(2)},$ $\chi^{(2)},$ and $u_{a}^{(2)}$ characterize the right-handed neutrino. At the same time, $\nu^{(1)},$ $\chi^{(1)},$ and $\nu_{a}^{(1)}$ respond to the right-handed antineutrino, and 
$\nu^{(2)},$ $\chi^{(2)},$ and $\nu_{a}^{(2)}$ correspond to the left-handed antineutrino.

It is already clear from the foregoing that the neutrino $\nu_{lL}$ and the antineutrino 
${\bar \nu}_{lR}$ refer to the left-polarized fermions, and the neutrino $\nu_{lR}$ and 
the antineutrino ${\bar \nu}_{lL}$ are of the right-polarized leptons. 

Such a full spin picture corresponding in an equation (\ref{8}) to the matrices (\ref{3})-(\ref{6}) and (\ref{11}) can be established by another way starting from (\ref{13}) if its prediction (\ref{62}) is carried out in nature.

At first sight, this says in favor of the compatibility of all requirements of a chiral invariance with implications of the helicity operator itself. On the other hand, such a unification of (\ref{12}) and (\ref{62}) shows that
\begin{equation}
m_{L}=-|{\bf p}_{L}|, \, \, \, \, m_{R}=|{\bf p}_{R}|, 
\label{86}
\end{equation}
and consequently, (\ref{13}) is one of those presentations of matrices $\alpha,$ $\beta,$ and 
$\gamma_{5},$ in each of which $\nu_{lR}$ and ${\bar \nu}_{lL}$ come forward as the particles, 
and $\nu_{lL}$ and ${\bar \nu}_{lR}$ are predicted as the antiparticles.

The difference in masses, energies, and momenta of a particle and an antiparticle violates, in the case of C-even types of leptons, their CPT-symmetry expressing the idea of a Lorentz invariance [21]. At the same time, a C-noninvariant neutrino itself regardless of whether or not an unbroken Lorentz symmetry exists in its nature, is strictly CPT-odd [22]. This does not imply of course that the same neutrino or antineutrino must be either fermion or antifermion. 

By following the structure of matrices (\ref{3})-(\ref{6}), (\ref{9}), and (\ref{11}),
it is easy to see that
\begin{equation}
[(\alpha{\bf p_{s}}+\beta m_{s}), \, \, \, \, \sigma{\bf p}_{s}]
=[\sigma{\bf p}_{s}, \, \, \, \, (\alpha{\bf p_{s}}+\beta m_{s})],
\label{87}
\end{equation}
\begin{equation}
[(\alpha{\bf p_{s}}+\beta m_{s}), \, \, \, \, \gamma_{5}] 
=[\gamma_{5}, \, \, \, \, (\alpha{\bf p_{s}}-\beta m_{s})],
\label{88}
\end{equation}
which characterize the behavior of the standard presentation (\ref{11}) both from the point 
of view of a C-even and from the point of view of a C-odd particles. 

To the same relationships (\ref{87}) and (\ref{88}) one can also lead by another way using (\ref{3}), (\ref{4}), (\ref{9}), and (\ref{13}), but the latter together with (\ref{5}), (\ref{6}), and (\ref{9}) satisfies the equalities
\begin{equation}
[(\alpha{\bf p_{s}}+\beta m_{s}), \, \, \, \, \sigma{\bf p}_{s}]
=[\sigma{\bf p}_{s}, \, \, \, \, (-\alpha{\bf p_{s}}+\beta m_{s})],
\label{89}
\end{equation}
\begin{equation}
[(\alpha{\bf p_{s}}+\beta m_{s}), \, \, \, \, \gamma_{5}]
=[\gamma_{5}, \, \, \, \, (-\alpha{\bf p_{s}}+\beta m_{s})].
\label{90}
\end{equation}

This would seem to say that either unification [18,22] of elementary objects in families of a different C-parity is incompatible with the chiral presentation (\ref{13}) or $\sigma{\bf p}_{s}$ 
is not helicity operator of a C-odd particle. On the other hand, as follows from symmetry laws, any C-invariant or C-noninvariant neutrino cannot simultaneously have both CPT-even vector and CPT-odd axial-vector nature. Such a circumstance becomes more interesting if we take into account that the existence of vector [23] and axial-vector [24,25] mirror Minkowski space-times are by no means excluded [8,9] experimentally. 

Thus, it follows that between the spontaneous mirror symmetry violation and the chiral presentation 
(\ref{13}) there exists a range of the structural contradictions, which expresses the ideas of the left- and right-handed neutrinos refering to long- and short-lived objects, respectively. These ideas require in principle to go away from the chiral definitions of the structure of matter fields taking into account that $\alpha,$ $\beta,$ and $\gamma_{5}$ come forward in (\ref{13}) as the matrices not connecting with the Dirac equation. Therefore, from the point of view of the mass, energy, and momentum matrices, each of (\ref{86}), (\ref{89}), and (\ref{90}) must be considered 
as an indication to the absence in a mirror world of a place for chirality.

There exist, however, a range of the structural equations, in each of which appears a part of a 
kind of system [26] of internally disclosed sets [27]. Exactly the same nature itself is not in force to exclude the existence of a kind of set criterion for matrices, for example, for matrices 
of a Dirac equation. Therefore, to establish the legality of each matrix from (\ref{11}) at the level of a very mirror space, one must follow the logic of the corresponding set of its elements.

\vspace{0.8cm}
\noindent
{\bf 6. Set criterion for matrices}
\vspace{0.4cm}

Our reasoning is such that each of the matrices
\begin{equation}
A=
\begin{pmatrix}
a_{11} & a_{12} & ... & a_{1n}\\
a_{21} & a_{22} & ... & a_{2n}\\
... & ... & ... & ...\\
a_{m1} & a_{m2} & ... & a_{mn}
\end{pmatrix},
\label{91}
\end{equation}
namely, of $m\times n$ matrices corresponds to a kind of set
$${\EuScript A}=\{a_{11}, \, \, \, \, a_{12}, \, \, \, \, ..., \, \, \, \, a_{1n},$$
\begin{equation}
a_{21}, \, \, \, \, a_{22}, \, \, \, \, ..., \, \, \, \, a_{2n}, \, \, \, \,
..., \, \, \, \, a_{m1}, \, \, \, \, a_{m2}, \, \, \, \, ..., \, \, \, \, a_{mn}\}.
\label{92}
\end{equation}

This is none other than a set-matrix duality principle, which expresses, for a matrix 
$\bigl(\, a_{11} \, \bigr)$ consisting only of one element $a_{11},$ the idea of a one-element set 
$\{\, a_{11} \, \},$ confirming that we cannot exclude the existence of an absolutely empty matrix 
$\bigl(\, \bigr)$ corresponding to a kind of empty set $\{\, \}$ in which there is no single element.

Furthermore, if $m=n=2,$ we have
\begin{equation}
A={{a_{11} \, \, \, \, \ a_{12}}\choose{a_{21} \, \, \, \, \ a_{22}}}.
\label{93}
\end{equation}

{\bf Lemma 10 (Theorem on the transitions between diagonals of a matrix).} There is no single transition between diagonals in a matrix without a permutation of its columns.

{\bf Proof of Lemma 10.} According to a set-matrix duality principle, the elements of such a matrix as (\ref{93}) constitute a quadratic set 
\begin{equation}
{\EuScript A}=\{a_{11}, \, \, \, \, a_{12}, \, \, \, \,
a_{21}, \, \, \, \, a_{22}\}.
\label{94}
\end{equation}

One of the admissible groupings of elements $a_{mn}$ transforms a set ${\EuScript A}$ from 
(\ref{94}) into 
\begin{equation}
{\EuScript A}=\{{\EuScript A_{1L}}, \, \, \, \, {\EuScript A_{2L}}, \, \, \, \, 
{\EuScript A_{1T}}, \, \, \, \, {\EuScript A_{2T}}, \, \, \, \, 
{\EuScript A_{1D}}, \, \, \, \, {\EuScript A_{2D}}\}
\label{95}
\end{equation}
in which
\begin{equation}
{\EuScript A_{1L}}=\{a_{11}, \, \, \, \, a_{12}\}, \, \, \, \,
{\EuScript A_{2L}}=\{a_{21}, \, \, \, \, a_{22}\},
\label{96}
\end{equation}
\begin{equation}
{\EuScript A_{1T}}=\{a_{11}, \, \, \, \, a_{21}\}, \, \, \, \,
{\EuScript A_{2T}}=\{a_{12}, \, \, \, \, a_{22}\},
\label{97}
\end{equation}
\begin{equation}
{\EuScript A_{1D}}=\{a_{11}, \, \, \, \, a_{22}\}, \, \, \, \,
{\EuScript A_{2D}}=\{a_{12}, \, \, \, \, a_{21}\}.
\label{98}
\end{equation}
Here ${\EuScript A_{1L}}$ and ${\EuScript A_{2L}}$ denote the first and second row or longitudinal 
$(L)$ sets of a matrix (\ref{93}), ${\EuScript A_{1T}}$ and ${\EuScript A_{2T}}$ describe its first and second column or transversal $(T)$ sets, ${\EuScript A_{1D}}$ and ${\EuScript A_{2D}}$ 
characterize in it the first and second diagonal $(D)$ sets. 

A feature of this structure is that we can replace in a set (\ref{95}) one longitudinal subset for another subset
\begin{equation}
{\EuScript A_{1L}}\rightarrow  {\EuScript A_{2L}} \, \, \Leftrightarrow \, \,
{\EuScript A_{2L}}\rightarrow {\EuScript A_{1L}}
\label{99}
\end{equation}
if and only if the very set-matrix duality principle admits a permutation of elements within each 
of the transversal and diagonal subsets
\begin{equation}
{\EuScript A_{1T}}=\{a_{21}, \, \, \, \, a_{11}\} \, \, \Leftrightarrow \, \,
{\EuScript A_{2T}}=\{a_{22}, \, \, \, \, a_{12}\},
\label{100}
\end{equation}
\begin{equation}
{\EuScript A_{1D}}=\{a_{22}, \, \, \, \, a_{11}\} \, \, \Leftrightarrow \, \,
{\EuScript A_{2D}}=\{a_{21}, \, \, \, \, a_{12}\}.
\label{101}
\end{equation}

They constitute for a matrix (\ref{93}) one more fully possible form
\begin{equation}
A={{a_{21} \, \, \, \, \ a_{22}}\choose{a_{11} \, \, \, \, \ a_{12}}}.
\label{102}
\end{equation}

However, we cannot replace one transversal subset for another subset
\begin{equation}
{\EuScript A_{1T}}\rightarrow  {\EuScript A_{2T}} \, \, \Leftrightarrow \, \,
{\EuScript A_{2T}}\rightarrow {\EuScript A_{1T}}
\label{103}
\end{equation}
until the very set-matrix duality principle is able to permutate both the
diagonal subsets
\begin{equation}
{\EuScript A_{1D}}\rightarrow  {\EuScript A_{2D}} \, \, \Leftrightarrow \, \,
{\EuScript A_{2D}}\rightarrow {\EuScript A_{1D}}
\label{104}
\end{equation} 
and the elements within each of the longitudinal subsets
\begin{equation}
{\EuScript A_{1L}}=\{a_{12}, \, \, \, \, a_{11}\} \, \, \Leftrightarrow \, \,
{\EuScript A_{2L}}=\{a_{22}, \, \, \, \, a_{21}\}.
\label{105}
\end{equation}

They establish for a matrix (\ref{93}) one more fully regular form
\begin{equation}
A={{a_{12} \, \, \, \, \ a_{11}}\choose{a_{22} \, \, \, \, \ a_{21}}}.
\label{106}
\end{equation}

This is exactly the same as when the transitions (\ref{104}) are carried out in a set (\ref{105}) 
as an indication in favor of each of (\ref{103}) and (\ref{105}), namely, in favor of Lemma 10 
and all consequences following from it. ${\square}$ 

If we choose $a_{11},$ $a_{12},$ $a_{21},$ and $a_{22}$ from the matrices (\ref{11}), any of (\ref{102}) and (\ref{106}) expresses in whole the idea of a kind of system of matrices from
\begin{equation}
\alpha={{\sigma \ \, \, \, \, \ 0}\choose{0 \ \, \, \, \, \ \sigma}}, \, \, \, \,
\beta={{0 \, \, -I}\choose{I \ \, \, \, \, \ 0}}, \, \, \, \,
\gamma_{5}={{I \ \, \, \, \, \ 0}\choose{0 \ \, \, \, \, \ I}},
\label{107}
\end{equation}
\begin{equation}
\alpha={{\sigma \ \, \, \, \, \ 0}\choose{0 \ \, \, \, \, \ \sigma}}, \, \, \, \,
\beta={{\ 0 \ \, \, \, \, \ I}\choose{-I \ \, \, \, \, \ 0}}, \, \, \, \,
\gamma_{5}={{I \ \, \, \, \, \ 0}\choose{0 \ \, \, \, \, \ I}},
\label{108}
\end{equation}
each of which is unlike (\ref{13}) a fully regular presentation of Dirac matrices. 

\vspace{0.8cm}
\noindent
{\bf 7. Internally disclosed matrices
of a Dirac equation}
\vspace{0.4cm}

To express the idea more clearly, it is desirable to use at first the Theorem 2, according to 
which, $m_{L},$ $E_{L},$ and ${\bf p}_{L}$ come forward as the objects of left points of a real number axis, and $m_{R},$ $E_{R},$ and ${\bf p}_{R}$ are predicted as the objects of its right points. They constitute herewith the internally disclosed sets such as 
\begin{equation}
\{m_{L}, \, \, \, \, m_{R}\}, \, \, \, \, \{E_{L}, \, \, \, \, E_{R}\}, \, \, \, \, 
\{{\bf p}_{L}, \, \, \, \, {\bf p}_{R}\}. 
\label{109}
\end{equation}

Each of them, from the point of view of Lemma 1, says about the existence of a kind of pair of fully regular one-element sets from its two elements
\begin{equation}
\{m_{L}\}, \, \, \, \, \{m_{R}\}, \, \, \, \, \{E_{L}\}, \, \, \, \, \{E_{R}\}, \, \, \, \,
\{{\bf p}_{L}\}, \, \, \, \, \{{\bf p}_{R}\}.
\label{110}
\end{equation}

If we now take into account that (\ref{109}) and (\ref{110}) are of sets from the diagonal elements  
of matrices (\ref{4}) and (\ref{6}), and consequently, all of the one-element sets 
\begin{equation}
\{m_{V}\}, \, \, \, \, \{E_{V}\}, \, \, \, \, \{{\bf p}_{V}\}, \, \, \, \, 
\{m_{A}\}, \, \, \, \, \{E_{A}\}, \, \, \, \, \{{\bf p}_{A}\} 
\label{111}
\end{equation}
have internal disclosure, then there arises a question of whether the absence of classification of matrices on the basis of a set-matrix duality principle is not strictly nonverisimilar even at the violation of some old rules of the very theory of their structure.

{\bf Definition 11.} A matrix is called fully regular or internally disclosed one if each set 
from its diagonal elements has an internal disclosure.

{\bf Definition 12.} A matrix is called fully casual or internally undisclosed one if each set 
from its diagonal elements has an internal undisclosure.

{\bf Lemma 11 (Theorem on the smallest nonempty matrix).} No single fully regular matrix from 
the same element exists without an internally disclosed matrix of a higher dimensionality.

{\bf Proof of Lemma 11.} The presence of an empty class [16] in all sets implies that in each matrix, an empty matrix not containing any element is necessarily present. This becomes possible owing to a set-matrix duality principle.

Thus, the existence of one of a set of one-element matrix element and a matrix of one-element set element is none other than a confirmation of the existence of both. ${\square}$

To the same assertion one can also lead by another way starting from the sets of the diagonal elements of matrices $\beta$ and $\gamma_{5},$ namely, from 
\begin{equation}
\{I\}, \, \, \, \, \{-I\}, \, \, \, \, \{I, \, \, \, \, -I\}, 
\label{112}
\end{equation}
for which a constancy law of the sum (\ref{48}) is not violated. 

Insofar as an internal disclosure of a set of elements of a matrix $\alpha$ is concerned, it will appear in the structure dependence of matrices (\ref{10}), because their diagonal elements 
constitute such sets as 
\begin{equation}
\{1\}, \, \, \, \, \{-1\}, \, \, \, \, \{i\}, \, \, \, \, \{-i\}, \, \, \, \, 
\{1, \, \, \, \, -1\}, \, \, \, \, \{i, \, \, \, \, -i\},
\label{113}
\end{equation}
which, according to Theorems 1 and 2, have internal disclosure. 

However, among the sets (\ref{109})-(\ref{113}) there are no sets mentioned in Lemmas 3, 5, 7, and 9, since in them appears a mathematical undisclosure of the same space, where they exist.

{\bf Axiom 3.} An internal disclosure of a matrix is none other than a mathematical disclosure 
of each pair of sets from its diagonal elements.
 
{\bf Axiom 4.} An internal undisclosure of a matrix is none other than a mathematical undisclosure of each pair of sets from its diagonal elements.

\vspace{0.8cm}
\noindent
{\bf 8. Matrix criterion for sets}
\vspace{0.4cm}

The preceding reasoning says that a matrix makes it possible to introduce the notions of the 
row, column, and diagonal of a set, confirming their availability in the defined order of elements.

{\bf Lemma 12 (Theorem on the diagonal elements of a set).} There is no single diagonal in 
an internally disclosed set without a crossing element of its row and column.

{\bf Proof of Lemma 12.} A set-matrix duality principle requires one to follow the logic, at the 
new level, of each matrix of (\ref{11}) from the point of view of his set. It chooses herewith 
the sets ${\EuScript A},$ ${\EuScript B},$ and $\varGamma_{5}$ for matrices $\alpha,$ $\beta,$ 
and $\gamma_{5}$ so that their structure had the form 
\begin{equation}
{\EuScript A}=\{0, \, \, \, \, \sigma, \, \, \, \, \sigma, \, \, \, \, 0\},
\label{114}
\end{equation}
\begin{equation}
{\EuScript B}=\{I, \, \, \, \, 0, \, \, \, \, 0, \, \, \, \, -I\},
\label{115}
\end{equation}
\begin{equation}
\varGamma_{5}=\{0, \, \, \, \, I, \, \, \, \, I, \, \, \, \, 0\}.
\label{116}
\end{equation}

To conform with a presentation (\ref{95}), the first of these sets involves the following longitudinal, transversal, and diagonal subsets: 
\begin{equation}
{\EuScript A_{1L}}=\{0, \, \, \, \, \sigma\}, \, \, \, \,
{\EuScript A_{2L}}=\{\sigma, \, \, \, \, 0\},
\label{117}
\end{equation}
\begin{equation}
{\EuScript A_{1T}}=\{0, \, \, \, \, \sigma\}, \, \, \, \,
{\EuScript A_{2T}}=\{\sigma, \, \, \, \, 0\},
\label{118}
\end{equation}
\begin{equation}
{\EuScript A_{1D}}=\{0, \, \, \, \, 0\}, \, \, \, \,
{\EuScript A_{2D}}=\{\sigma, \, \, \, \, \sigma\}.
\label{119}
\end{equation}

One can also present a set
\begin{equation}
{\EuScript B}=\{{\EuScript B_{1L}}, \, \, \, \, {\EuScript B_{2L}}, \, \, \, \, 
{\EuScript B_{1T}}, \, \, \, \, {\EuScript B_{2T}}, \, \, \, \, 
{\EuScript B_{1D}}, \, \, \, \, {\EuScript B_{2D}}\}
\label{120}
\end{equation}
with longitudinal, transversal, and diagonal subsets
\begin{equation}
{\EuScript B_{1L}}=\{I, \, \, \, \, 0\}, \, \, \, \,
{\EuScript B_{2L}}=\{0, \, \, \, \, -I\},
\label{121}
\end{equation}
\begin{equation}
{\EuScript B_{1T}}=\{I, \, \, \, \, 0\}, \, \, \, \,
{\EuScript B_{2T}}=\{0, \, \, \, \, -I\},
\label{122}
\end{equation}
\begin{equation}
{\EuScript B_{1D}}=\{I, \, \, \, \, -I\}, \, \, \, \,
{\EuScript B_{2D}}=\{0, \, \, \, \, 0\}.
\label{123}
\end{equation}

As well as in (\ref{120}), each subset of a set  
\begin{equation}
\varGamma_{5}=\{\varGamma_{5}^{1L}, \, \, \, \, \varGamma_{5}^{2L}, \, \, \, \, 
\varGamma_{5}^{1T}, \, \, \, \, \varGamma_{5}^{2T}, \, \, \, \, 
\varGamma_{5}^{1D}, \, \, \, \, \varGamma_{5}^{2D}\}
\label{124}
\end{equation}
has a self compound structure
\begin{equation}
\varGamma_{5}^{1L}=\{0, \, \, \, \, I\}, \, \, \, \,
\varGamma_{5}^{2L}=\{I, \, \, \, \, 0\},
\label{125}
\end{equation}
\begin{equation}
\varGamma_{5}^{1T}=\{0, \, \, \, \, I\}, \, \, \, \,
\varGamma_{5}^{2T}=\{I, \, \, \, \, 0\},
\label{126}
\end{equation}
\begin{equation}
\varGamma_{5}^{1D}=\{0, \, \, \, \, 0\}, \, \, \, \,
\varGamma_{5}^{2D}=\{I, \, \, \, \, I\}.
\label{127}
\end{equation}

They together with subets of sets (\ref{114}) and (\ref{120}) constitute the nine pairs of 
subsets from two elements such that their structure recognizes the existence in each of sets 
${\EuScript A},$ ${\EuScript B},$ and $\varGamma_{5}$ of one pair of diagonals with the crossing elements of its row and column, thereby confirming the validity of Lemma 12 and all of its implications. ${\square}$  

Based on the matrices (\ref{3}), one can also find that 
\begin{equation}
{\EuScript M_{s}}=\{m_{V}, \, \, \, \, 0, \, \, \, \, 0, \, \, \, \, m_{V}\},
\label{128}
\end{equation}
\begin{equation}
{\EuScript E_{s}}=\{E_{V}, \, \, \, \, 0, \, \, \, \, 0, \, \, \, \, E_{V}\},
\label{129}
\end{equation}
\begin{equation}
{\EuScript P_{s}}=\{{\bf p}_{V}, \, \, \, \, 0, \, \, \, \, 0, \, \, \, \, {\bf p}_{V}\}.
\label{130}
\end{equation}

The structural set ${\EuScript M_{s}}$ consisting of elements of a matrix $m_{s}$ can be presented 
by the following manner:
\begin{equation}
{\EuScript M_{s}}=\{{\EuScript M_{s}^{1L}}, \, \, \, \, {\EuScript M_{s}^{2L}}, \, \, \, \, 
{\EuScript M_{s}^{1T}}, \, \, \, \, {\EuScript M_{s}^{2T}}, \, \, \, \, 
{\EuScript M_{s}^{1D}}, \, \, \, \, {\EuScript M_{s}^{2D}}\}.
\label{131}
\end{equation}

Its longitudinal, transversal, and diagonal subsets have the form
\begin{equation}
{\EuScript M_{s}^{1L}}=\{m_{V}, \, \, \, \, 0\}, \, \, \, \,
{\EuScript M_{s}^{2L}}=\{0, \, \, \, \, m_{V}\},
\label{132}
\end{equation}
\begin{equation}
{\EuScript M_{s}^{1T}}=\{m_{V}, \, \, \, \, 0\}, \, \, \, \,
{\EuScript M_{s}^{2T}}=\{0, \, \, \, \, m_{V}\},
\label{133}
\end{equation}
\begin{equation}
{\EuScript M_{s}^{1D}}=\{m_{V}, \, \, \, \, m_{V}\}, \, \, \, \,
{\EuScript M_{s}^{2D}}=\{0, \, \, \, \, 0\}.
\label{134}
\end{equation}

The elements of a matrix $E_{s}$ lead to the substitution of a set ${\EuScript E_{s}}$ for
\begin{equation}
{\EuScript E_{s}}=\{{\EuScript E_{s}^{1L}}, \, \, \, \, {\EuScript E_{s}^{2L}}, \, \, \, \, 
{\EuScript E_{s}^{1T}}, \, \, \, \, {\EuScript E_{s}^{2T}}, \, \, \, \, 
{\EuScript E_{s}^{1D}}, \, \, \, \, {\EuScript E_{s}^{2D}}\},
\label{135}
\end{equation}
which consists of the longitudinal, transversal, and diagonal subsets 
\begin{equation}
{\EuScript E_{s}^{1L}}=\{E_{V}, \, \, \, \, 0\}, \, \, \, \,
{\EuScript E_{s}^{2L}}=\{0, \, \, \, \, E_{V}\},
\label{136}
\end{equation}
\begin{equation}
{\EuScript E_{s}^{1T}}=\{E_{V}, \, \, \, \, 0\}, \, \, \, \,
{\EuScript E_{s}^{2T}}=\{0, \, \, \, \, E_{V}\},
\label{137}
\end{equation}
\begin{equation}
{\EuScript E_{s}^{1D}}=\{E_{V}, \, \, \, \, E_{V}\}, \, \, \, \,
{\EuScript E_{s}^{2D}}=\{0, \, \, \, \, 0\}.
\label{138}
\end{equation}

Finally, for a set
\begin{equation}
{\EuScript P_{s}}=\{{\EuScript P_{s}^{1L}}, \, \, \, \, 
{\EuScript P_{s}^{2L}}, \, \, \, \, 
{\EuScript P_{s}^{1T}}, \, \, \, \, {\EuScript P_{s}^{2T}}, \, \, \, \, 
{\EuScript P_{s}^{1D}}, \, \, \, \, {\EuScript P_{s}^{2D}}\},
\label{139}
\end{equation}
including the elements of a matrix ${\bf p}_{s},$ the structural subets of this class 
can have the following structure:
\begin{equation}
{\EuScript P_{s}^{1L}}=\{{\bf p}_{V}, \, \, \, \, 0\}, \, \, \, \,
{\EuScript P_{s}^{2L}}=\{0, \, \, \, \, {\bf p}_{V}\},
\label{140}
\end{equation}
\begin{equation}
{\EuScript P_{s}^{1T}}=\{{\bf p}_{V}, \, \, \, \, 0\}, \, \, \, \,
{\EuScript P_{s}^{2T}}=\{0, \, \, \, \, {\bf p}_{V}\},
\label{141}
\end{equation}
\begin{equation}
{\EuScript P_{s}^{1D}}=\{{\bf p}_{V}, \, \, \, \, {\bf p}_{V}\}, \, \, \, \,
{\EuScript P_{s}^{2D}}=\{0, \, \, \, \, 0\}.
\label{142}
\end{equation}

So, we must recognize that 
\begin{equation}
{\EuScript A}=\{{\EuScript A_{1L}}, \, \, \, \, {\EuScript A_{2L}}, \, \, \, \, 
{\EuScript A_{1T}}, \, \, \, \, {\EuScript A_{2T}}\},
\label{143}
\end{equation}
\begin{equation}
{\EuScript B}=\{{\EuScript B_{1L}}, \, \, \, \, {\EuScript B_{2L}}, \, \, \, \, 
{\EuScript B_{1T}}, \, \, \, \, {\EuScript B_{2T}}\},
\label{144}
\end{equation}
\begin{equation}
\varGamma_{5}=\{\varGamma_{5}^{1L}, \, \, \, \, \varGamma_{5}^{2L}, \, \, \, \, 
\varGamma_{5}^{1T}, \, \, \, \, \varGamma_{5}^{2T}\},
\label{145}
\end{equation}
\begin{equation}
{\EuScript M_{s}}=\{{\EuScript M_{s}^{1L}}, \, \, \, \, {\EuScript M_{s}^{2L}}, \, \, \, \, 
{\EuScript M_{s}^{1T}}, \, \, \, \, {\EuScript M_{s}^{2T}}\},
\label{146}
\end{equation}
\begin{equation}
{\EuScript E_{s}}=\{{\EuScript E_{s}^{1L}}, \, \, \, \, {\EuScript E_{s}^{2L}}, \, \, \, \, 
{\EuScript E_{s}^{1T}}, \, \, \, \, {\EuScript E_{s}^{2T}}\},
\label{147}
\end{equation}
\begin{equation}
{\EuScript P_{s}}=\{{\EuScript P_{s}^{1L}}, \, \, \, \, 
{\EuScript P_{s}^{2L}}, \, \, \, \, 
{\EuScript P_{s}^{1T}}, \, \, \, \, {\EuScript P_{s}^{2T}}\}
\label{148}
\end{equation}
are fully regular sets of a Dirac equation
\begin{equation}
{\EuScript E_{s}}\psi_{s}={\EuScript H_{s}}\psi_{s}
\label{149}  
\end{equation} 
such that in it
\begin{equation}
{\EuScript H_{s}}={\EuScript A}{\EuScript P_{s}}+{\EuScript B}{\EuScript M_{s}},
\label{150}
\end{equation}
\begin{equation}
\psi_{s}={\EuScript U_{s}}({\bf p}_{s}, \sigma)e^{-ip_{s} \cdot{\it x}_{s}}, \, \, \, \, E_{s}>0,
\label{1151}
\end{equation}
\begin{equation}
{\EuScript U_{s}}={\EuScript U^{(r)}}=\{\chi^{(r)}, \, \, \, \, u_{a}^{(r)}\},
\label{152}
\end{equation}
\begin{equation}
{\EuScript U_{s}}=\{{\EuScript U_{s}^{1L}}, \, \, \, \, 
{\EuScript U_{s}^{2L}}, \, \, \, \, {\EuScript U_{s}^{1T}}\},
\label{153}
\end{equation}
\begin{equation}
{\EuScript U_{s}^{1L}}=\{\chi^{(r)}\}, \, \, \, \,
{\EuScript U_{s}^{2L}}=\{u_{a}^{(r)}\}, \, \, \, \,
{\EuScript U_{s}^{1T}}=\{\chi^{(r)}, \, \, \, \, u_{a}^{(r)}\},
\label{154}
\end{equation}
\begin{equation}
\psi_{s}={\EuScript N_{s}}({\bf p}_{s}, \sigma)e^{-ip_{s} \cdot{\it x}_{s}}, \, \, \, \, E_{s}<0,
\label{155}
\end{equation}
\begin{equation}
{\EuScript N_{s}}={\EuScript N^{(r)}}=\{\nu_{a}^{(r)}, \, \, \, \, \chi^{(r)}\},
\label{156}
\end{equation}
\begin{equation}
{\EuScript N_{s}}=\{{\EuScript N_{s}^{1L}}, \, \, \, \, 
{\EuScript N_{s}^{2L}}, \, \, \, \, {\EuScript N_{s}^{1T}}\},
\label{157}
\end{equation}
\begin{equation}
{\EuScript N_{s}^{1L}}=\{\nu_{a}^{(r)}\}, \, \, \, \,
{\EuScript N_{s}^{2L}}=\{\chi^{(r)}\}, \, \, \, \,
{\EuScript N_{s}^{1T}}=\{\nu_{a}^{(r)}, \, \, \, \, \chi^{(r)}\}.
\label{158}
\end{equation}

Here ${\EuScript U_{s}^{1L}}$ and ${\EuScript U_{s}^{2L}}$ express the first and second row of 
a set ${\EuScript U_{s}}$ of a matrix (\ref{53}), ${\EuScript U_{s}^{1T}}$ characterizes in it a single column, ${\EuScript N_{s}^{1L}}$ and ${\EuScript N_{s}^{2L}}$ describe the first and second row of a set ${\EuScript N_{s}}$ of a matrix (\ref{64}), ${\EuScript N_{s}^{1T}}$ implies a singleness of its column. 

If we now use the sets (\ref{143}) and (\ref{148}), we see that   
$${\EuScript A}{\EuScript P_{s}}=
\{{\EuScript A_{1L}}, \, \, \, \, {\EuScript A_{2L}}, \, \, \, \, 
{\EuScript A_{1T}}, \, \, \, \, {\EuScript A_{2T}}\} 
\{{\EuScript P_{s}^{1L}}, \, \, \, \, 
{\EuScript P_{s}^{2L}}, \, \, \, \, 
{\EuScript P_{s}^{1T}}, \, \, \, \, {\EuScript P_{s}^{2T}}\}=$$
$$\{{\EuScript A_{1L}}{\EuScript P_{s}^{1T}}, \, \, \, \,
{\EuScript A_{1L}}{\EuScript P_{s}^{2T}}, \, \, \, \,
{\EuScript A_{2L}}{\EuScript P_{s}^{1T}}, \, \, \, \,
{\EuScript A_{2L}}{\EuScript P_{s}^{2T}}\}=$$
$$\{\{0, \, \, \, \, \sigma\}
\{{\bf p}_{V}, \, \, \, \, 0\}, \, \, \, \,
\{0, \, \, \, \, \sigma\}
\{0, \, \, \, \, {\bf p}_{V}\},$$
$$\{\sigma, \, \, \, \, 0\}
\{{\bf p}_{V}, \, \, \, \, 0\}, \, \, \, \,
\{\sigma, \, \, \, \, 0\}
\{0, \, \, \, \, {\bf p}_{V}\}\}=$$
$$\{0 \cdot{\bf p}_{V}+\sigma \cdot0, \, \, \, \,
0 \cdot0+\sigma \cdot{\bf p}_{V},$$
$$\sigma \cdot{\bf p}_{V}+0 \cdot0, \, \, \, \,
\sigma \cdot0+0 \cdot{\bf p}_{V}\}=$$
\begin{equation}
\{0, \, \, \, \,
\sigma{\bf p}_{V}, \, \, \, \,
\sigma{\bf p}_{V}, \, \, \, \,
0\}.
\label{159}
\end{equation}

Performing the same set operations that led to (\ref{159}), but having in mind the sets 
${\EuScript B}$ and ${\EuScript M_{s}},$ we found
$${\EuScript B}{\EuScript M_{s}}=
\{{\EuScript B_{1L}}, \, \, \, \, {\EuScript B_{2L}}, \, \, \, \, 
{\EuScript B_{1T}}, \, \, \, \, {\EuScript B_{2T}}\}
\{{\EuScript M_{s}^{1L}}, \, \, \, \, {\EuScript M_{s}^{2L}}, \, \, \, \, 
{\EuScript M_{s}^{1T}}, \, \, \, \, {\EuScript M_{s}^{2T}}\}=$$
$$\{{\EuScript B_{1L}}{\EuScript M_{s}^{1T}}, \, \, \, \,
{\EuScript B_{1L}}{\EuScript M_{s}^{2T}}, \, \, \, \,
{\EuScript B_{2L}}{\EuScript M_{s}^{1T}}, \, \, \, \,
{\EuScript B_{2L}}{\EuScript M_{s}^{2T}}\}=$$
$$\{\{I, \, \, \, \, 0\}\{m_{V}, \, \, \, \, 0\}, \, \, \, \,
\{I, \, \, \, \, 0\}\{0, \, \, \, \, m_{V}\},$$
$$\{0, \, \, \, \, -I\}\{m_{V}, \, \, \, \, 0\}, \, \, \, \,
\{0, \, \, \, \, -I\}\{0, \, \, \, \, m_{V}\}\}=$$
$$\{I \cdot m_{V}+0 \cdot0, \, \, \, \,
I \cdot0+0 \cdot m_{V},$$
$$0 \cdot m_{V}+(-I) \cdot0, \, \, \, \,
0 \cdot0+(-I) \cdot m_{V}\}=$$
\begin{equation}
\{m_{V}, \, \, \, \,
0, \, \, \, \,
0, \, \, \, \,
-m_{V}\}.
\label{160}
\end{equation}

At these situations, (\ref{159}) and (\ref{160}) replace (\ref{150}) for
$${\EuScript H_{s}}={\EuScript A}{\EuScript P_{s}}+{\EuScript B}{\EuScript M_{s}}=$$
$$\{0, \, \, \, \, \sigma{\bf p}_{V}, \, \, \, \, \sigma{\bf p}_{V}, \, \, \, \, 0\}+
\{m_{V}, \, \, \, \, 0, \, \, \, \, 0, \, \, \, \, -m_{V}\}=$$
$$\{0+m_{V}, \, \, \, \, \sigma{\bf p}_{V}+0, \, \, \, \, 
\sigma{\bf p}_{V}+0, \, \, \, \, 0-m_{V}\}=$$
\begin{equation}
\{m_{V}, \, \, \, \, \sigma{\bf p}_{V}, \, \, \, \, 
\sigma{\bf p}_{V}, \, \, \, \, -m_{V}\}
\label{161}
\end{equation}
and thereby confirm that a set 
\begin{equation}
{\EuScript H_{s}}=\{{\EuScript H_{s}^{1L}}, \, \, \, \, {\EuScript H_{s}^{2L}}, \, \, \, \, 
{\EuScript H_{s}^{1T}}, \, \, \, \, {\EuScript H_{s}^{2T}}\}
\label{162}
\end{equation}
involves both the row $(L)$ and column $(T)$ subsets:
\begin{equation}
{\EuScript H_{s}^{1L}}=\{m_{V}, \, \, \, \, \sigma{\bf p}_{V}\}, \, \, \, \,
{\EuScript H_{s}^{2L}}=\{\sigma{\bf p}_{V}, \, \, \, \, -m_{V}\},
\label{163}
\end{equation}
\begin{equation}
{\EuScript H_{s}^{1T}}=\{m_{V}, \, \, \, \, \sigma{\bf p}_{V}\}, \, \, \, \,
{\EuScript H_{s}^{2T}}=\{\sigma{\bf p}_{V}, \, \, \, \, -m_{V}\}.
\label{164}
\end{equation}

Thus, it follows that 
$${\EuScript H_{s}}{\EuScript U_{s}}=\{{\EuScript H_{s}^{1L}}, \, \, \, \, 
{\EuScript H_{s}^{2L}}, \, \, \, \, {\EuScript H_{s}^{1T}}, \, \, \, \, 
{\EuScript H_{s}^{2T}}\}$$
$$\{{\EuScript U_{s}^{1L}}, \, \, \, \, 
{\EuScript U_{s}^{2L}}, \, \, \, \, {\EuScript U_{s}^{1T}}\}=
\{{\EuScript H_{s}^{1L}}{\EuScript U_{s}^{1T}}, \, \, \, \,
{\EuScript H_{s}^{2L}}{\EuScript U_{s}^{1T}}\}=$$
$$\{\{m_{V}, \, \, \, \, \sigma{\bf p}_{V}\}\{\chi^{(r)}, \, \, \, \, u_{a}^{(r)}\}, \, \, \, \,
\{\sigma{\bf p}_{V}, \, \, \, \, -m_{V}\}\{\chi^{(r)}, \, \, \, \, u_{a}^{(r)}\}\}=$$
$$\{m_{V} \cdot \chi^{(r)}+\sigma{\bf p}_{V} \cdot u_{a}^{(r)}, \, \, \, \,
\sigma{\bf p}_{V} \cdot\chi^{(r)}+(-m_{V}) \cdot u_{a}^{(r)}\}=$$
\begin{equation}
\{m_{V}\chi^{(r)}+(\sigma{\bf p}_{V})u_{a}^{(r)}, \, \, \, \,
(\sigma{\bf p}_{V})\chi^{(r)}-m_{V}u_{a}^{(r)}\},
\label{165}
\end{equation}
and consequently, the multiplication ${\EuScript E_{s}}{\EuScript U_{s}}$ is equal to
$${\EuScript E_{s}}{\EuScript U_{s}}=\{{\EuScript E_{s}^{1L}}, \, \, \, \, 
{\EuScript E_{s}^{2L}}, \, \, \, \, {\EuScript E_{s}^{1T}}, \, \, \, \, 
{\EuScript E_{s}^{2T}}\}$$
$$\{{\EuScript U_{s}^{1L}}, \, \, \, \, 
{\EuScript U_{s}^{2L}}, \, \, \, \, {\EuScript U_{s}^{1T}}\}=
\{{\EuScript E_{s}^{1L}}{\EuScript U_{s}^{1T}}, \, \, \, \,
{\EuScript E_{s}^{2L}}{\EuScript U_{s}^{1T}}\}=$$
$$\{\{E_{V}, \, \, \, \, 0\}\{\chi^{(r)}, \, \, \, \, u_{a}^{(r)}\}, \, \, \, \,
\{0, \, \, \, \, E_{V}\}\{\chi^{(r)}, \, \, \, \, u_{a}^{(r)}\}\}=$$
\begin{equation}
\{E_{V} \cdot \chi^{(r)}+0 \cdot u_{a}^{(r)}, \, \, \, \,
0 \cdot\chi^{(r)}+E_{V} \cdot u_{a}^{(r)}\}=
\{E_{V}\chi^{(r)}, \, \, \, \, E_{V}u_{a}^{(r)}\}.
\label{166}
\end{equation}

They show [8] that a system
\begin{equation}
E_{V}\chi^{(r)}=(\sigma{\bf p}_{V})u_{a}^{(r)}+m_{V}\chi^{(r)},
\label{167}
\end{equation}
\begin{equation}
E_{V}u_{a}^{(r)}=(\sigma{\bf p}_{V})\chi^{(r)}-m_{V}u_{a}^{(r)}
\label{168}
\end{equation}
is none other than an equality
\begin{equation}
\{E_{V}\chi^{(r)}, \, \, \, \, E_{V}u_{a}^{(r)}\}=
\{m_{V}\chi^{(r)}+(\sigma{\bf p}_{V})u_{a}^{(r)}, \, \, \, \,
(\sigma{\bf p}_{V})\chi^{(r)}-m_{V}u_{a}^{(r)}\}.
\label{169}
\end{equation}

In a similar way, one can define ${\EuScript E_{s}}{\EuScript N_{s}}$ and
${\EuScript H_{s}}{\EuScript N_{s}}$ and find [8] the equations
\begin{equation}
|E_{V}|\nu_{a}^{(r)}=-(\sigma{\bf p}_{V})\chi^{(r)}-m_{V}\nu_{a}^{(r)},
\label{170}
\end{equation}
\begin{equation}
|E_{V}|\chi^{(r)}=-(\sigma{\bf p}_{V})\nu_{a}^{(r)}+m_{V}\chi^{(r)}
\label{171}
\end{equation}
from an equality
\begin{equation}
\{|E_{V}|\nu_{a}^{(r)}, \, \, \, \, |E_{V}|\chi^{(r)}\}=
\{-(\sigma{\bf p}_{V})\chi^{(r)}-m_{V}\nu_{a}^{(r)}, \, \, \, \,
m_{V}\chi^{(r)}-(\sigma{\bf p}_{V})\nu_{a}^{(r)}\}.
\label{172}
\end{equation}

For completeness, we remark that the matrices (\ref{5}) suggest
\begin{equation}
{\EuScript M_{s}}=\{0, \, \, \, \, m_{A}, \, \, \, \, m_{A}, \, \, \, \, 0\},
\label{173}
\end{equation}
\begin{equation}
{\EuScript E_{s}}=\{0, \, \, \, \, E_{A}, \, \, \, \, E_{A}, \, \, \, \, 0\},
\label{174}
\end{equation}
\begin{equation}
{\EuScript P_{s}}=\{0, \, \, \, \, {\bf p}_{A}, \, \, \, \, {\bf p}_{A}, \, \, \, \, 0\}.
\label{175}
\end{equation}

Therefore, if one starts from (\ref{173})-(\ref{175}) and then performs explicit set operations
of a Dirac equation, one can establish that 
\begin{equation}
\{E_{A}u_{a}^{(r)}, \, \, \, \, E_{A}\chi^{(r)}\}=
\{(\sigma{\bf p}_{A})\chi^{(r)}+m_{A}u_{a}^{(r)}, \, \, \, \,
(\sigma{\bf p}_{A})u_{a}^{(r)}-m_{A}\chi^{(r)}\}
\label{176}
\end{equation}
holds if there exist [9] the united connections
\begin{equation}
E_{A}u_{a}^{(r)}=(\sigma{\bf p}_{A})\chi^{(r)}+m_{A}u_{a}^{(r)},
\label{177}
\end{equation}
\begin{equation}
E_{A}\chi^{(r)}=(\sigma{\bf p}_{A})u_{a}^{(r)}-m_{A}\chi^{(r)}.
\label{178}
\end{equation}

As in ratios (\ref{125})-(\ref{127}), the equality
\begin{equation}
\{|E_{A}|\chi^{(r)}, \, \, \, \, |E_{A}|\nu_{a}^{(r)}\}=
\{-(\sigma{\bf p}_{A})\nu_{a}^{(r)}-m_{A}\chi^{(r)}, \, \, \, \, 
-(\sigma{\bf p}_{A})\chi^{(r)}+m_{A}\nu_{a}^{(r)}\},
\label{179}
\end{equation}
found in a given case leads to a system 
\begin{equation}
|E_{A}|\chi^{(r)}=-(\sigma{\bf p}_{A})\nu_{a}^{(r)}-m_{A}\chi^{(r)},
\label{180}
\end{equation}
\begin{equation}
|E_{A}|\nu_{a}^{(r)}=-(\sigma{\bf p}_{A})\chi^{(r)}+m_{A}\nu_{a}^{(r)}
\label{181}
\end{equation}
as to one more consequence of a set-matrix duality principle. 

\vspace{0.8cm}
\noindent
{\bf 9. Concluding remarks}
\vspace{0.4cm}

Furthermore, if it turns out that there is no single internally disclosed set, for which 
a fully regular matrix would not exist, we cannot exclude the possibility to strictly change 
our presentations about set operations. Without such a change, the mathematically united theory construction of sets and matrices still remains not quite in line with logic.

Therefore, to have the right to perform explicit set operations necessary for deciding the problems at the new level, we must at first define the rows, columns, and diagonals of each of them if this is not forbidden by quantities of its elements. 

For such purposes, it is desirable to use here the following sets:
\begin{equation}
{\EuScript C}=\{\varepsilon\},
\label{182}
\end{equation}
\begin{equation}
{\EuScript D}=\{\eta, \, \, \, \, \kappa\},
\label{183}
\end{equation}
\begin{equation}
{\EuScript E}=\{\lambda, \, \, \, \, \mu, \, \, \, \, \nu\},
\label{184}
\end{equation}
\begin{equation}
{\EuScript F}=\{\rho, \, \, \, \, \tau, \, \, \, \,
\upsilon, \, \, \, \, \omega\},
\label{185}
\end{equation}
\begin{equation}
{\EuScript H}=\{\phi, \, \, \, \, \psi, \, \, \, \,
\varphi, \, \, \, \, \chi\}.
\label{186}
\end{equation}

A set ${\EuScript C}$ consists only of one element. There is neither a row nor a column in 
a one-elemet set. The existence of either one row ${\EuScript C_{L}}$ or one pair of rows
${\EuScript C_{1L}}$ and ${\EuScript C_{2L}}$ of the same column ${\EuScript C_{T}}$ for 
${\EuScript C}$ would indicate that it is not one-elemet. However, according to Lemma 11, a matrix 
$\bigl(\, \varepsilon \, \bigr)$ of a set $\{\varepsilon\}$ exists in a kind of fully regular matrix 
as one of its nonempty submatrix. 

A two-elemet set ${\EuScript D}$ must distinguish itself from ${\EuScript C}$ by either one 
row or two rows of its one column. Therefore, it should be defined the matrix structure of 
a set ${\EuScript D}$ so that its elements have constituted in whole either a set of the row 
${\EuScript D_{L}}$ such as (\ref{183}) or two nonempty rows ${\EuScript D_{1L}}$ and 
${\EuScript D_{2L}}$ of the same nonempty column ${\EuScript D_{1T}}$ of a set of the column 
${\EuScript D_{T}}$ in the form
\begin{equation}
{\EuScript D_{L}}={\EuScript D}=\{\eta, \, \, \, \, \kappa\},
\label{187}
\end{equation}
\begin{equation}
{\EuScript D_{T}}={\EuScript D}=\{{\EuScript D_{1L}}, \, \, \, \, 
{\EuScript D_{2L}}, \, \, \, \, {\EuScript D_{1T}}\},
\label{188}
\end{equation}
\begin{equation}
{\EuScript D_{1L}}=\{\eta\}, \, \, \, \,
{\EuScript D_{2L}}=\{\kappa\}, \, \, \, \,
{\EuScript D_{1T}}=\{\eta, \, \, \, \, \kappa\}.
\label{189}
\end{equation}

On this basis, each matrix of (\ref{53}) and (\ref{64}) itself has constituted a kind of set of the column from (\ref{153}) and (\ref{157}) as a consequence of a set-matrix duality principle. 

The possible grouping of elements of a set ${\EuScript E}$ would seem to say about that it 
has two rows ${\EuScript E_{1L}}$ and ${\EuScript E_{2L}}$ from two elements and one column 
${\EuScript D_{1T}}$ from three elements. This implication, however, does not correspond to 
reality. The point is that the availability of such subsets in ${\EuScript E}$ is incompatible 
with an equality of the number of its rows and the number of elements of its column. In other 
words, there is no single row with two elements in a set ${\EuScript E}$ even if it is strictly 
a set of the row. A set ${\EuScript E}$ can therefore be presented either as a set of the row 
${\EuScript E_{L}}$ or as a set of the column ${\EuScript E_{T}}$ by the following manner:
\begin{equation}
{\EuScript E_{L}}={\EuScript E}=\{\lambda, \, \, \, \, \mu, \, \, \, \, \nu\},
\label{190}
\end{equation}
\begin{equation}
{\EuScript E_{T}}={\EuScript E}=\{{\EuScript E_{1L}}, \, \, \, \, 
{\EuScript E_{2L}}, \, \, \, \, {\EuScript E_{3L}}, \, \, \, \,{\EuScript E_{1T}}\},
\label{191}
\end{equation}
\begin{equation}
{\EuScript E_{1L}}=\{\lambda\}, \, \, \, \,
{\EuScript E_{2L}}=\{\mu\}, \, \, \, \,
{\EuScript E_{3L}}=\{\nu\}, \, \, \, \, 
{\EuScript E_{1T}}=\{\lambda, \, \, \, \, \mu, \, \, \, \, \nu\}.
\label{192}
\end{equation}

One more characteristic moment is an equal number of rows ${\EuScript X_{L}}$ and columns
${\EuScript X_{T}}$ of a quadratic set ${\EuScript X_{LT}}$ from
$${\EuScript X}={\EuScript C}, \, \, \, \, {\EuScript D}, \, \, \, \, 
{\EuScript E}, \, \, \, \, {\EuScript F}, \, \, \, \, {\EuScript H}, ....$$ 

{\bf Definition 13.} An Internally disclosed set is called a quadratic one if it has the same quantity of rows and columns.

Such a correspondence expresses, for each of sets ${\EuScript F}$ and ${\EuScript H},$ the idea 
of a kind of pair of diagonals ${\EuScript X_{1D}}$ and ${\EuScript X_{2D}},$ allowing one to 
define its self matrix structure
\begin{equation}
{\EuScript F_{LT}}={\EuScript F}=\{{\EuScript F_{1L}}, \, \, \, \, {\EuScript F_{2L}}, \, \, \, \, 
{\EuScript F_{1T}}, \, \, \, \, {\EuScript F_{2T}}, \, \, \, \, 
{\EuScript F_{1D}}, \, \, \, \, {\EuScript F_{2D}}\},
\label{193}
\end{equation}
\begin{equation}
{\EuScript F_{1L}}=\{\rho, \, \, \, \, \tau\}, \, \, \, \,
{\EuScript F_{2L}}=\{\upsilon, \, \, \, \, \omega\},
\label{194}
\end{equation}
\begin{equation}
{\EuScript F_{1T}}=\{\rho, \, \, \, \, \upsilon\}, \, \, \, \,
{\EuScript F_{2T}}=\{\tau, \, \, \, \, \omega\},
\label{195}
\end{equation}
\begin{equation}
{\EuScript F_{1D}}=\{\rho, \, \, \, \, \omega\}, \, \, \, \,
{\EuScript F_{2D}}=\{\tau, \, \, \, \, \upsilon\},
\label{196}
\end{equation}
\begin{equation}
{\EuScript H_{LT}}={\EuScript H}=\{{\EuScript H_{1L}}, \, \, \, \, {\EuScript H_{2L}}, \, \, \, \, 
{\EuScript H_{1T}}, \, \, \, \, {\EuScript H_{2T}}, \, \, \, \, 
{\EuScript H_{1D}}, \, \, \, \, {\EuScript H_{2D}}\},
\label{197}
\end{equation}
\begin{equation}
{\EuScript H_{1L}}=\{\phi, \, \, \, \, \psi\}, \, \, \, \,
{\EuScript H_{2L}}=\{\varphi, \, \, \, \, \chi\},
\label{198}
\end{equation}
\begin{equation}
{\EuScript H_{1T}}=\{\phi, \, \, \, \, \varphi\}, \, \, \, \,
{\EuScript H_{2T}}=\{\psi, \, \, \, \, \chi\},
\label{199}
\end{equation}
\begin{equation}
{\EuScript H_{1D}}=\{\phi, \, \, \, \, \chi\}, \, \, \, \,
{\EuScript H_{2D}}=\{\psi, \, \, \, \, \varphi\},
\label{200}
\end{equation}
which was used above. 

An intramatrix feature of the two types of sets, namely, the sets of the row ${\EuScript X_{L}}$ 
and the column ${\EuScript X_{T}},$ is their simultaneous absence, their coexistence or both. In 
the first case, the matrix has a set consisting of one element. The matrices with sets of the same quantity $(m=n)$ of rows $(mL)$ and columns $(nT)$ refer to the second case. An example of the 
third case is the matrix with a set with elements of either one ${\EuScript X_{L}}$ row or one 
${\EuScript X_{T}}$ column. 

{\bf Definition 14.} An Internally disclosed set is called a set of the row if and only if each object of each point constitutes in whole a kind of empty row of its column.

{\bf Definition 15.} An Internally disclosed set is called a set of the column if and only if each object of each point constitutes in whole a kind of nonempty row of its column.

We will now perform, on the basis of a set-matrix duality principle, the explicit set operations
referring to those sets, the matrix structure of which were already established above. This requires one to explain, for example, the equality of sets ${\EuScript F}$ and ${\EuScript H}$ as a consequence of equalities
\begin{equation}
{\EuScript F_{mL}}={\EuScript H_{mL}} \, \, \Leftrightarrow \, \,  
{\EuScript F_{nT}}={\EuScript H_{nT}},
\label{201}
\end{equation}
expressing at $m=n=1,2$ the idea about that
\begin{equation}
\rho=\phi, \, \, \, \, \tau=\psi, \, \, \, \,
\upsilon=\varphi, \, \, \, \, \omega=\chi.
\label{202}
\end{equation}

Insofar as their inequality is concerned, it is in favor of 
\begin{equation}
{\EuScript F_{mL}}\neq {\EuScript H_{mL}} \, \, \Leftrightarrow \, \,  
{\EuScript F_{nT}}\neq {\EuScript H_{nT}},
\label{203}
\end{equation}
which at $m=n=1,2$ state that 
\begin{equation}
\rho\neq \phi, \, \, \, \, \tau\neq \psi, \, \, \, \,
\upsilon\neq \varphi, \, \, \, \, \omega\neq \chi.
\label{204}
\end{equation}

Uniting ${\EuScript F}$ with ${\EuScript H}$ having (\ref{193}), (\ref{194}), (\ref{197}), and (\ref{198}), one can found that
$${\EuScript K}={\EuScript F}+{\EuScript H}=$$
$$\{{\EuScript F_{1L}}, \, \, \, \, {\EuScript F_{2L}}\}+
\{{\EuScript H_{1L}}, \, \, \, \, {\EuScript H_{2L}}\}=$$
\begin{equation}
\{\rho+\phi, \, \, \, \, \tau+\psi, \, \, \, \, 
\upsilon+\varphi, \, \, \, \, \omega+\chi\}.
\label{205}
\end{equation}

Unification of this type is none other than a set 
\begin{equation}
{\EuScript K_{LT}}={\EuScript K}=\{{\EuScript K_{1L}}, \, \, \, \, {\EuScript K_{2L}}, \, \, \, \, 
{\EuScript K_{1T}}, \, \, \, \, {\EuScript K_{2T}}, \, \, \, \, 
{\EuScript K_{1D}}, \, \, \, \, {\EuScript K_{2D}}\},
\label{206}
\end{equation}
\begin{equation}
{\EuScript K_{1L}}=\{\rho+\phi, \, \, \, \, \tau+\psi\}, \, \, \, \,
{\EuScript K_{2L}}=\{\upsilon+\varphi, \, \, \, \, \omega+\chi\},
\label{207}
\end{equation}
\begin{equation}
{\EuScript K_{1T}}=\{\rho+\phi, \, \, \, \, \upsilon+\varphi\}, \, \, \, \,
{\EuScript K_{2T}}=\{\tau+\psi, \, \, \, \, \omega+\chi\},
\label{208}
\end{equation}
\begin{equation}
{\EuScript K_{1D}}=\{\rho+\phi, \, \, \, \, \omega+\chi\}, \, \, \, \,
{\EuScript K_{2D}}=\{\tau+\psi, \, \, \, \, \upsilon+\varphi\}.
\label{209}
\end{equation}

One more consequence of the matrix structure of sets ${\EuScript F}$ and ${\EuScript H}$
is such that their multiplication constitutes in whole another set 
$${\EuScript M}={\EuScript F}{\EuScript H}=
\{{\EuScript F_{1L}}, \, \, \, \, {\EuScript F_{2L}}, \, \, \, \, 
{\EuScript F_{1T}}, \, \, \, \, {\EuScript F_{2T}}\} 
\{{\EuScript H_{1L}}, \, \, \, \, {\EuScript H_{{2L}}}, \, \, \, \, 
{\EuScript H_{1T}}, \, \, \, \, {\EuScript H_{2T}}\}=$$
$$\{{\EuScript F_{1L}}{\EuScript H_{1T}}, \, \, \, \,
{\EuScript F_{1L}}{\EuScript H_{2T}}, \, \, \, \,
{\EuScript F_{2L}}{\EuScript H_{1T}}, \, \, \, \,
{\EuScript F_{2L}}{\EuScript H_{2T}}\}=$$
$$\{\{\rho, \, \, \, \, \tau\}\{\phi, \, \, \, \, \varphi\}, \, \, \, \,
\{\rho, \, \, \, \, \tau\}\{\psi, \, \, \, \, \chi\},$$
$$\{\upsilon, \, \, \, \, \omega\}\{\phi, \, \, \, \, \varphi\}, \, \, \, \,
\{\upsilon, \, \, \, \, \omega\}\{\psi, \, \, \, \, \chi\}\}=$$
$$\{\rho \cdot\phi+\tau \cdot\varphi, \, \, \, \,
\rho \cdot\psi+\tau \cdot\chi,$$
$$\upsilon \cdot\phi+\omega \cdot\varphi, \, \, \, \,
\upsilon \cdot\psi+\omega \cdot\chi\}=$$
\begin{equation}
\{\rho\phi+\tau\varphi, \, \, \, \, \rho\psi+\tau\chi, \, \, \, \,
\upsilon\phi+\omega\varphi, \, \, \, \, \upsilon\psi+\omega\chi\}
\label{210}
\end{equation}
and that, consequently, we have
\begin{equation}
{\EuScript M_{LT}}={\EuScript M}=\{{\EuScript M_{1L}}, \, \, \, \, {\EuScript M_{2L}}, \, \, \, \, 
{\EuScript M_{1T}}, \, \, \, \, {\EuScript M_{2T}}, \, \, \, \, 
{\EuScript M_{1D}}, \, \, \, \, {\EuScript M_{2D}}\},
\label{211}
\end{equation}
\begin{equation}
{\EuScript M_{1L}}=\{\rho\phi+\tau\varphi, \, \, \, \, \rho\psi+\tau\chi\}, \, \, \, \,
{\EuScript M_{2L}}=\{\upsilon\phi+\omega\varphi, \, \, \, \, \upsilon\psi+\omega\chi\},
\label{212}
\end{equation}
\begin{equation}
{\EuScript M_{1T}}=\{\rho\phi+\tau\varphi, \, \, \, \, \upsilon\phi+\omega\varphi\}, \, \, \, \,
{\EuScript M_{2T}}=\{\rho\psi+\tau\chi, \, \, \, \, \upsilon\psi+\omega\chi\},
\label{213}
\end{equation}
\begin{equation}
{\EuScript M_{1D}}=\{\rho\phi+\tau\varphi, \, \, \, \, \upsilon\psi+\omega\chi\}, \, \, \, \,
{\EuScript M_{2D}}=\{\rho\psi+\tau\chi, \, \, \, \, \upsilon\phi+\omega\varphi\}.
\label{214}
\end{equation}

If we now suppose that ${\EuScript K}={\EuScript M},$ the united connections 
\begin{equation}
{\EuScript K_{mL}}={\EuScript M_{mL}} \, \, \Leftrightarrow \, \, 
{\EuScript K_{nT}}={\EuScript M_{nT}}
\label{215}
\end{equation}
suggest at $m=n=1,2$ the same system
\begin{equation}
\rho+\phi=\rho\phi+\tau\varphi, \, \, \, \, 
\tau+\psi=\rho\psi+\tau\chi, 
\label{216}
\end{equation}
\begin{equation}
\upsilon+\varphi=\upsilon\phi+\omega\varphi, \, \, \, \, 
\omega+\chi=\upsilon\psi+\omega\chi.
\label{217}
\end{equation}

However, the very coexistence of a matrix of set elements and a set of matrix elements indicates 
the role of the unified system of rows, columns, and diagonals in all set operations. A notion 
about matrix determinant may therefore be connected with a notion about set determinant. Such a nonclassical correspondence, regardless of what is the interaction between the matrix and its set, requires one to solve the question as to what is the value of determinants of the discussed types 
of quadratic sets. For this we must at first introduce a symbol $\varDelta_{\EuScript X}$ for its denotation, allowing us to write at ${\EuScript X}={\EuScript F},$ ${\EuScript H}$ the following determinants of the second order: 
\begin{equation}
\varDelta_{\EuScript F}=\{{\EuScript F_{1D}}, \, \, \, \, {\EuScript F_{2D}}\}=
\rho\omega-\tau\upsilon,
\label{218}
\end{equation}
\begin{equation}
\varDelta_{\EuScript H}=\{{\EuScript H_{1D}}, \, \, \, \, {\EuScript H_{2D}}\}=
\phi\chi-\psi\varphi.
\label{219}
\end{equation}

{\bf Axiom 5.} An element of the smallest nonempty matrix is none other than a determinant 
of a kind of quadratic set. 

{\bf Axiom 6.} An element of the smallest nonempty set is none other than a determinant 
of a kind of quadratic matrix. 

Thus, we can use, for example, each of real (imaginary) numbers as a determinant of either a kind 
of quadratic set or a kind of quadratic matrix such that it comes forward in a real (an imaginary) number axis as an object of one of its real (imaginary) points.

Finally, for future, the operations over a matrix of a higher dimensionality and over a set of a higher cardinality together with some implications of a set-matrix duality principle (not noted here) will be presented in our further works. 

\vspace{0.8cm}
\noindent
{\bf Acknowledgment}
\vspace{0.3cm}

I would like to express my gratitude to Shavkat A. Alimov and Karimbergen K. Kudaybergenov for 
their numerous valuable discussions and critical remarks, which were useful in improving the manuscript.

\vspace{0.8cm}
\noindent
{\bf References}
\begin{enumerate}
\item
Sharafiddinov RS. Vector currents of massive neutrinos of an electroweak nature. 
{\it Canadian Journal of Physics.} 2014; 92: 262-1268.

Available from: https://doi.org//10.1139/cjp-2013-0458.
\item
Glashow SL. Partial symmetries of weak interactions. {\it Nuclear Physics B.}
1961; 22: 579-588.
\item
Salam A, Ward JC. Gauge theory of elementary interactions. {\it Physics Letters B.}
1964; 136: 763-768.
\item
Weinberg S. A model of leptons. {\it Physical Review Letters.}
1967; 19: 1264-1266.

\item
Sharafiddinov RS. A new family with a fourth lepton flavor.  
{\it Physics Essays.} 2017; 30: 150-155. Available from: https://doi.org/10.4006/0836-1398-30.2.150.
\item 
Zel'dovich YaB, Perelomov AB. An influence of the weak interaction on particle electromagnetic properties. {\it Journal of Experimental and Theoretical Physics.} 1960; 39: 1115-1125.
\item 
Sharafiddinov RS. On a latent structure of lepton universality. {\it The Europian Physical Journal Plus.} 2011; 126: 40.

Available from: https://doi.org/10.1140/epjp/i2011-11040-x.
\item 
Sharafiddinov RS. Nature itself in a mirror space-time. {\it Canadian Journal of Physics.}
2015; 93: 1005-1008.
Available from: https://doi.org//10.1139/cjp-2014-0497.
\item 
Sharafiddinov RS. An unbroken axial-vector current conservation law. {\it International Journal 
of Theoretical Physics.} 2016; 55: 2139-2147.

Available from: https://doi.org/10.1007/s10773-015-2852-3.
\item 
Dirac PAM. The quantum theory of electron. {\it Proceedings of the Royal Society of London A.} 
1928; 117: 610-624.
Available from: https://doi.org/10.1098/rspa.1928.0023. 
\item 
Weyl H. Elektron und gravitation. {\it Magazine for Physics.} 1929; 56: 330–352. 

Available from: https://doi.org/10.1007/BF01339504. 

\item 
Sharafiddinov RS. Unification theorems of elements of a set. Abstracts of Papers Presented to the American Mathematical Society. 2020; 41 (2): 1158-03-99.

\item 
Sharafiddinov RS. Fully regular sets of an imaginary space. {\it Contemporary Mathematics.} 
2023; 4 (4): 817-829. 
Available from: https://doi.org/10.37256/cm.4420232405.
\item 
Sharafiddinov RS. Internally disclosed sets of a real space. {\it Contemporary Mathematics.} 
2024; 5 (2): 1655-1670.
Available from: https://doi.org/10.37256/cm.5220243023.
\item 
Alexandroff PS. {\it Introduction to the Theory of Sets and the General Topology.} 
Moscow: Nauka; 1977.
\item 
Wells W, Yart WW. {\it First Year Algebra.} Boston: D. C. Heath \& Company; 1912. 
\item 
Dedekind R. {\it Stetigkeit und Irrationale Zahlen.} Braunschweig: Friedrich Vieweg; 1872.

\item 
Sharafiddinov RS. True neutrality as a new type of flavor. {\it International Journal 
of Theoretical Physics.} 2016; 55: 3041–3058.

Available from: https://doi.org/10.1007/s10773-016-2936-8.
\item 
Sharafiddinov RS. On a mass-charge structure of gauge invariance.
{\it Physics Essays.} 2016; 29: 410-415.
Available from: https://doi.org/10.4006/0836-1398-29.3.410. 
\item 
Sharafiddinov RS. An axial-vector photon in a mirror world. {\it Contemporary Mathematics.} 
2024; 5 (4): 5328-5340.
Available from: https://doi.org/10.37256/cm.5420242746.
\item  
Greenberg OW. CPT violation implies violation of Lorentz invariance.
{\it Physical Review Letters.}
2002; 89: 231602.

Available from: https://doi.org/10.1103/PhysRevLett.89.231602.
\item 
Sharafiddinov RS. A theory of flavor and set of the interaction structural parts. {\it Journal of Physical and Natural Sciences.} 2013; 4: 1-11. 

Available from: https://doi.org/10.48550/arXiv.physics/0702233.
\item 
Adamson P, Anghel1 I, Backhouse G, Barr G, Bisha M, Blake A, et al.
Measurement of neutrino and antineutrino oscillations using beam and atmospheric data in MINOS.
{\it Physical Review Letters.}
2013; 110: 251801. 

Available from: https://doi.org/10.1103/PhysRevLett.110.251801.
\item 
Kuzmin KS, Lyubushkin VV, Naumov VA. Quasielastic axial-vector mass from experiments on 
neutrino-nucleus scattering. {\it The Europian Journal of Physics C.} 2008; 54: 517-538.

Available from: https://doi.org/10.1140/epjc/s10052-008-0582-x.
\item 
Gran G, Jeon EJ, Aliu E, Andringa S, Aoki S, Argyriades J, Asakura K, Ashie R, Berghaus F, et al. Measurement of the quasielastic axial vector mass in neutrino interactions on oxygen. 
{\it Physical Review D.} 2006; 74: 052002.

Available from: https://doi.org/10.1103/PhysRevD.74.052002.
\item 
Sharafiddinov RS. An allgravity as a grand unification of forces. {\it Physics Essays.} 
2021; 34(3): 398-410. 
Available from: https://doi.org/10.4006/0836-1398-34.3.397.

\item 
Sharafiddinov RS. Set criterion for selected quantum theory equations.  
Abstracts of Papers Presented to the American Mathematical Society. 2020; 41 (2): 1158-81-100. 

\end{enumerate}

\end{document}